\documentclass[letterpaper,12pt,openright]{book}
\usepackage[latin1]{inputenc}
\usepackage{fancyhdr}
\fancypagestyle{plain}{%
\fancyhead{} 
}
\usepackage{amsfonts}
\usepackage{amssymb}
\usepackage{graphicx}
\usepackage{setspace}
\usepackage[mathcal]{euscript}
\usepackage{latexsym}
\newcommand{\bea}{\begin{eqnarray}}
\newcommand{\eea}{\end{eqnarray}}
\newcommand{\be}{\begin{equation}}
\newcommand{\ee}{\end{equation}}
\newcommand{\bdm}{\begin{displaymath}}
\newcommand{\edm}{\end{displaymath}}

\newcommand{\rd}{\mathrm{d}}

\begin{document}
\renewcommand{\listfigurename}{Lista de Figuras}
\renewcommand{\contentsname}{\'Indice}
\renewcommand{\listtablename}{Lista de Tablas}
\renewcommand{\bibname}{Bibliograf\'{\i}a}
\renewcommand{\indexname}{Indice alfab\'etico}
\renewcommand{\figurename}{Figura}
\renewcommand{\tablename}{Tabla}
\renewcommand{\partname}{Parte}
\renewcommand{\chaptername}{Cap\'{\i}tulo}
\renewcommand{\appendixname}{Ap\'endice}
\begin{center}
\thispagestyle{empty}
Universidad Central de Venezuela\\
Facultad de Ciencias\\
Escuela de F\'{\i}sica\\
\vspace{5cm} \textbf{Dualidad en teor\'{\i}as de spin 2 masivo en
dimensi\'on 2+1}\\
\vspace{3cm}
Dr. P\'{\i}o J. Arias\\
\vspace{5 cm}
Trabajo de ascenso presentado\\
ante la ilustre Universidad Central de Venezuela\\
para optar a la categor\'{\i}a de\\
{\bf Profesor Asociado}\\
\vspace{2cm}
 Caracas, 11 de febrero de 2005
\end{center}
\newpage
\thispagestyle{empty}
\begin{center}
\vspace{3cm} {\bf Resumen\\
\vspace{1cm} Dualidad en teor\'{\i}as de spin 2 masivo en
dimensi\'on 2+1}\\
\vspace{1cm}
Dr. P\'{\i}o J. Arias \\
\vspace{1cm}
Universidad Central de Venezuela \\
\vspace{1cm}
\end{center}
{\narrower
 Se obtienen las ecuaciones que deben satisfacer los distintos campos que describen realizaciones
 del \'algebra del grupo de Poincar\'e en dimensi\'on 2+1, con masa m y spin entero (no cero). Para
 el caso de spin 2 se muestra que hay tres modelos posibles cuyas acciones se pueden conectar por
 transformaciones de dualidad. Estas transformaciones incorporan las invariancias de calibre que
 diferencian un modelo de otro. Se discute brevemmente la relaci\'on entre las funciones
 de partici\'on de estos modelos}

\newpage
\thispagestyle{empty} \vspace{10cm}
\begin{flushright}
{\it{A P\'{\i}o Jos\'e, Dafne y Zulay}}
\end{flushright}
\newpage
\thispagestyle{empty}
\newpage
\pagenumbering{roman}

\tableofcontents
\newpage
\pagenumbering{arabic}
\setcounter{page}{1}

\chapter{Introducci\'on}

El estudio de teor\'{\i}as vectoriales y tensoriales en
dimensi\'on 2+1 estuvo, inicialmente motivado por su conexi\'on
con el comportamiento de modelos, en 3+1 dimensiones, a altas
temperaturas \cite{gross}. Sin embargo, en el correr del tiempo la
f\'isica en esta dimensi\'on espacio-temporal ha adquirido
importancia propia dado que ha servido para aportar nuevas ideas a
la f\'isica en 3+1 dimensiones.

Es en esta dimensi\'on donde se introduce el t\'ermino de
teori\'ias masivas topol\'ogicas, las cuales tienen la interesante
propiedad de no explotar la invariancia de calibre \cite{DJT}.
Estos modelos masivos tienen en com\'un que son sensibles a las
transformaciones discretas de inversi\'on temporal y de paridad,
en concordancia con las particularidades que deben tener las
representaciones del \'algrebra del grupo de Poincar\'e en esta
dimensi\'on.

Tambi\'en es propio de la teor\'ia de campos en 2+1 dimensiones la
aparici\'on de los modelos masivos autoduales para teor\'ias con
spin 1, 2, 3 y 4 \cite{PK,DJself}\cite{ArKh,deserself}\cite{adel}.
En estos modelos la ecuaci\'on de movimiento relaciona al
''potencial'' con la ''intensidad de campo'', y corresponde a la
realizaci\'on que debe satisfacer el casimir
$\mathbb{P}.\mathbb{J}$ del grupo de Poincar\'e, con $\mathbb{P}$
la realizaci\'on del momentum lineal y $\mathbb{J}$ la
realizaci\'on de la parte de spin del dual del momentum angular
\begin{equation}
\Psi=\frac{1}{ms}(\mathbb{P}.\mathbb{J})\Psi, \label{ecusd}
\end{equation}
donde $\Psi$ es un objeto que transforma linealmente bajo el grupo
de Lorentz. Para $s=1,2$ la ecuaci\'on (\ref{ecusd}) se escribe en
componentes, respectivamente, como
\begin{eqnarray}
a_{\mu}&=&\frac{1}{m}f_{\mu}(a),\\
{h_{\mu}^{Tt}}^a&=&\frac{1}{m}{\omega_{\mu}}^a(h^{Tt}),
\end{eqnarray}
donde
$f_{\mu}(a)={\varepsilon_{\mu}}^{\nu\lambda}\partial_{\nu}a_{\lambda}$
corresponde al dual de la intensidad del campo de Maxwell y
${\omega_{\mu}}^a(h)$ corresponde al dual de la conexi\'on de spin
sin torsi\'on, linealizada (${h_{\mu}^{Tt}}^a$ es sim\'etrico,
transverso y sin traza). \'Estas ecuaciones pueden escribirse para
otros objetos y corresponden a otras teor\'ias de spin 1 y 2
equivalentes a las autoduales, y que a su vez gozan de tener
invariancias de calibre adicionales. En el caso de $s=1$ la
ecuaci\'on tambi\'en corresponde a una realizaci\'on del \'algebra
de Poincar\'e si tomamos como campo matriz al dual del campo de
Maxwell $F_{\mu\nu}(a)$. La teor\'ia correspondiente es el modelo
topol\'ogico masivo, para el cual se conoce bastante que bien que
es dual a la teor\'ia masiva autodual vectorial y resulta
completamente equivalente si el espacio base es topol\'ogicamente
trivial.

Para la ecuaci\'on autodual de spin 2, puede verse que \'esta
tambi\'en se realiza si en lugar del ${h_{\mu}}^a$ consideramos a
la conexi\'on de spin linealizada ${\omega_{\mu}}^a$ o al tensor
de Einstein Linealizado ${G_{\mu}}^a$. En el primer caso
corresponde al modelo intermedio \cite{ArKh} y en el segundo la
model topol\'ogico masivo linealizado \cite{DJT}. Estos tres
modelos tienen el mismo espectro y se ha visto que podemos a
partir del modelo topol\'ogico masivo hasta el autodual fijando
lasa simetr\'ias de calibre que posee el modelo original
\cite{jorgepio}.

En este trabajo veremos que, tal como sucede en los modelos
vectoriales, los modelos de spin 2 estan conectados por
transformaciones de dualidad que van incorporando las invariancias
de calibre que distinguen a un modelo de otro. En el Cap\'itulo 2
haremos una r\'apida introducci\'on al grupo de Poincar\'e y sus
representaciones, obteniendo as\'i de, manera heur\'istica, las
ecuaciones de movimieto que deben cumplir las realizaciones
masivas del \'algebra para spin entero (no nulo). Luego en el
Cap\'itulo 3 veremos como es la situaci\'on entre los modelos
vectoriales, como se conectan por dualidad y se asoman algunos
resultados concernientes a la realci\'on entre las funciones de
partici\'on de los modelos considerados. En el Cap\'itulo
siguiente se repite el programa para los tres modelos de spin 2.

Las notaciones y convenciones para las teor\'ias de spin 2 se dan
en el Ap\'endice.

\chapter{Ecuaciones de movimiento para part\'{\i}culas masivas con spin entero en $D=2+1$}

En \'este cap\'itulo hacemos un pasaje r\'apido por el grupo de
Poincar\'e siguiendo lineamientos convencionales~\cite{weinberg}.
Introduciendo las particularidades que surgen en dimensi\'on $2+1$
veremos de manera heur\'istica cuales son las ecuaciones de
movimiento movimiento que deben cumplir las distintas
representaciones del grupo con masa y spin entero generalizando
los resultados ya conocidos para spines bajos~\cite{jackiwnair}.
La presentaci\'on no pretende ser rigurosa, sino mas bien de una
forma que se vea como aparecen las ecuaciones de movimiento que
deben satisfacer los campos que representan part\'iculas con masa
y spin entero. En los cap\'itulos siguientes veremos que estas
corresponder\'an a distintos modelos que estan conectados por
transformaciones de dualidad en los casos de $s=1,2$.

\section{El grupo de Poincar\'e y sus generadores}

El grupo de Poincar\'e, tambi\'en llamado de Lorentz
inhomog\'eneo, est\'a corformado por las transformaciones que
dejan invariante el intervalo $\rd x^{\mu}\rd x_{\mu}$.

La ley de transformaci\'on es

\begin{equation}
x^{\mu}\longrightarrow x^{'\mu}={\Lambda^{\mu}}_{\nu}x^{\nu} +
c^{\mu},
\end{equation}
con $\Lambda^{\mu}_{\nu}$ una matriz que preserva la m\'etrica de
Minkowski (una matriz del grupo de Lorentz, $O(3,1)$) que
satisface

\begin{equation}
\eta_{\mu\nu}={\Lambda^{\lambda}}_{\mu}{\Lambda^{\rho}}_{\nu}\eta_{\lambda\rho},
\end{equation}
donde $\eta_{\mu\nu}=diag(-1,+1,+1,)$. $c^{\mu}$ es un vector
constante, que genera una traslaci\'on en el espacio tiempo.

Si representamos los elementos del grupo de Poincar\'e por
$(\mathbf{\Lambda},c)$, $\mathbf{\Lambda}$ la matriz de Lorentz y
$c$ el vector de traslaci\'on, el producto de dos elementos del
grupo se representa por:

\begin{equation}\label{eq:ap13}
(\mathbf{\Lambda}_{2},c_{2})(\mathbf{\Lambda}_{1},c_{1})=
(\mathbf{\Lambda}_{2}\mathbf{\Lambda}_{1},\mathbf{\Lambda}_{2}c_{1}
+c_{2}).
\end{equation}
Al cuantizar, estas transformaciones estar\'an representadas por
operadores unitarios lineales o antiunitarios antilineales de
forma de preservar la probabilidad de transici\'on entre
estados~\cite{weinberg}. As\'i

\begin{eqnarray}
U(\mathbf{\Lambda},c) & = & U(\mathbb{I},c)U(\mathbf{\Lambda},0),\nonumber \\
& \equiv & U(c)U(\mathbf{\Lambda}).
\end{eqnarray}
Si $U(c)=e^{-ic_{\mu}\mathbb{P}^{\mu}}$, con $\mathbb{P}^{\mu}$
herm\'itico. Tendremos que

\begin{equation}
 U(c^{'})U(c)=U(c +c^{'}),
\end{equation}

de donde

\begin{equation}
[\mathbb{P}^{\mu},\mathbb{P}^{\nu}]=0.
\end{equation}
Si tomamos
$U(\mathbf{\Lambda})=e^{\frac{i}{2}\omega_{\mu\nu}\mathbb{M}^{\mu\nu}}$,
entonces de (\ref{eq:ap13})

$$ U(c')U(\mathbf{\Lambda'})U(c)U(\mathbf{\Lambda})=U(c')U(\mathbf{\Lambda'}c')
U(\mathbf{\Lambda'})U(\mathbf{\Lambda}),$$

\'o

\begin{equation}
U(\mathbf{\Lambda'})U(c)U^{-1}(\mathbf{\Lambda'})=U(\mathbf{\Lambda'}c),
\end{equation}
de donde, al desarrollar las exponenciales, llegamos a que
$\mathbb{P}^{\mu}$ transforma como un vector

\begin{equation}
U(\mathbf{\Lambda})\mathbb{P}^{\mu}U^{-1}(\mathbf{\Lambda})=
{(\mathbf{\Lambda}^{-1})^{\mu}}_{\nu}\mathbb{P}^{\nu}
\end{equation}
y adicionalmente, teniendo en cuenta que la transformaci\'on es
infinitesimal,i.e.\\
${\mathbf{\Lambda}^{\mu}}_{\nu}={\delta^{\mu}}_{\nu}+{\omega^{\mu}}_{\nu}$

\begin{equation}
i[\mathbb{P}^{\alpha},\mathbb{M}^{\mu\nu}]=
\eta^{\mu\alpha}\mathbb{P}^{\nu} -
\eta^{\nu\alpha}\mathbb{P}^{\mu}.
\end{equation}

Finalmente, del hecho que
$U^{-1}(\mathbf{\Lambda})=U(\mathbf{\Lambda}^{-1})$, tendremos que

\begin{equation}
U(\mathbf{\Lambda'})U(\mathbf{\Lambda})U^{-1}(\mathbf{\Lambda'})=
U(\mathbf{\Lambda'}\mathbf{\Lambda}\mathbf{\Lambda'}^{-1}),
\end{equation}
de donde se obtiene las reglas de conmutaci\'on del operador
herm\'itico $\mathbb{M}^{\mu\nu}$

\begin{eqnarray}
i[\mathbb{M}^{\mu\nu},\mathbb{M}^{\rho\sigma}]=
\eta^{\mu\rho}\mathbb{M}^{\nu\sigma} - \eta^{\nu\rho}\mathbb{M}^{\mu\sigma}
-\eta^{\mu\sigma}\mathbb{M}^{\nu\rho} +
\eta^{\nu\sigma}\mathbb{M}^{\mu\rho}.
\end{eqnarray}

En $D=2+1$, tenemos la particularidad de que como
$\mathbb{M}^{\mu\nu}=-\mathbb{M}^{\nu\mu}$ podemos introducir su
dual

\begin{equation}
\mathbb{J}^{\mu}=\frac{1}{2}\varepsilon^{\mu\nu\lambda}\mathbb{M}_{\nu\lambda},
\end{equation}
y, entonces, las relaciones de conmutaci\'on entre los generadores
quedan como

\begin{eqnarray}\label{eq:ap113}
i[\mathbb{P}^{\mu},\mathbb{P}^{\nu}] & = & 0 \ \ \ \ \ \ \ \ \ \ \ \ \ \ \  \textrm{(a)}, \nonumber \\
i[\mathbb{P}^{\mu},\mathbb{J}^{\nu}] & = &
-\varepsilon^{\mu\nu\lambda}\mathbb{P}_{\lambda}\ \ \ \ \ \textrm{(b)},\\
i[\mathbb{J}^{\mu},\mathbb{J}^{\nu}] & = &
-\varepsilon^{\mu\nu\lambda}\mathbb{J}_{\lambda}\ \ \ \ \
\textrm{(c)},\nonumber
\end{eqnarray}
que se refiere como el \'algebra entre los generadores del grupo
de  Poincar\'e. Es f\'acil ver que $\mathbb{J}^{0}$ esta asociada
a las rotaciones en el plano ($U(R)=e^{i\theta\mathbb{J}^{0}}$,
para $\omega_{ij}=-\varepsilon_{ij}\theta$) y $\mathbb{J}^{i}$ a
los ``boost''.

Otra particularidad interesante de $D=2+1$ es la consideraci\'on
de las transformaciones discretas impropias. En el caso de la
inversi\'on temporal la transformaci\'on es la usual

\begin{eqnarray}
&T\nonumber \\
&(x^{0},x^{1},x^{2}) \longrightarrow (-x^{0},x^{1},x^{2}),
\end{eqnarray}
Sin embargo la inversi\'on espacial $(x^{0},x^{1},x^{2})
\longrightarrow (x^{0},-x^{1},-x^{2})$ corresponde a una
rotaci\'on de $180^{o}$, que es una transformaci\'on propia, es
decir $det\mathbf{\Lambda}_{IE}=+1$. En esta dimensi\'on la
transformaci\'on que se refiere como paridad, es la
transformaci\'on impropia

\begin{eqnarray}
&P \nonumber \\
&(x^{0},x^{1},x^{2}) \longrightarrow (x^{0},-x^{1},x^{2}).
\end{eqnarray}

Sean $\mathcal{P}^{\nu}_{\mu}$ y $\mathcal{T}^{\nu}_{\mu}$ la
representaci\'on matricial de las transformaciones de inversi\'on
temporal y paridad Los operadores que las representan ser\'an

\begin{equation}
T=U(\mathcal{T},0) \quad, \quad P=U(\mathcal{P},0).
\end{equation}
As\'i, para una transformaci\'on $(\Lambda,c)$ sucede que

\begin{eqnarray}\label{eq:ap117}
PU(\Lambda,c)P^{-1} & = &
U(\mathcal{P}\Lambda\mathcal{P}^{-1},\mathcal{P}c)\ \ \ \ \ \textrm{(a)}, \nonumber \\
TU(\Lambda,c)T^{-1} & = &
T(\mathcal{T}\Lambda\mathcal{T}^{-1},\mathcal{T}c)\ \ \ \ \
\textrm{(b)},
\end{eqnarray}
con $U(\Lambda,c)= \mathbb{I} -ic_{\mu}\mathbb{P}^{\mu} +
\frac{i}{2}\omega_{\mu\nu}\mathbb{M}^{\mu\nu} + \cdots$.

Desarrollando miembro a miembro en (\ref{eq:ap117}) obtenemos

\begin{eqnarray}\label{eq:ap118}
Pi\mathbb{P}^{\mu}P^{-1} & =
& i{\mathcal{P}^{\mu}}_{\nu}\mathbb{P}^{\nu},\ \ \ \ \ \ \ \ \ \ \ \ \ \textrm{(a)} \nonumber \\
Pi\mathbb{M}^{\mu\nu}P^{-1} & =
& i{\mathcal{P}^{\mu}}_{\alpha}{\mathcal{P}^{\nu}}
_{\beta}\mathbb{M}^{\alpha\beta},\ \ \ \ \ \textrm{(b)}\\
Ti\mathbb{P}^{\mu}T^{1} & =
& i{\mathcal{T}^{\mu}}_{\nu}\mathbb{P}^{\nu},\ \ \ \ \ \ \ \ \ \ \ \ \  \textrm{(c)} \nonumber \\
Ti\mathbb{M}^{\mu\nu}T^{-1} & = &
i{\mathcal{T}^{\mu}}_{\alpha}{\mathcal{T}^{\nu}}_{\beta}\mathbb{M}^{\alpha\beta},\
\ \ \ \ \textrm{(d)}\nonumber
\end{eqnarray}
donde se han dejado los $i$ expl\'icitos dado que no sabemos si
los operadores $P $ o $T$ son unitarios y lineales o antiunitarios
y antilineales. Dado que $P^{o}\sim H$ (la energ\'ia), si tomamos
$\mu=0$ en (\ref{eq:ap118}a) notamos que si supieramos que $P$ es
antiunitario y antilineal llegariamos a una situaci\'on
f\'isicamente inaceptable. As\'i $P$ es unitario y lineal.
An\'analogamente tomando  $\mu=0$ en (\ref{eq:ap118}c),
obtendremos que para tener una transformaci\'on f\'isicamente
aceptable requerimos que $T$ sea antiunitario y antilineal. Por
tanto (\ref{eq:ap118}) se escribe como

\begin{eqnarray}\label{eq:ap119}
P\mathbb{P}^{\mu}P^{-1} & = &
{\mathcal{P}^{\mu}}_{\nu}\mathbb{P}^{\nu},\ \ \ \ \ \ \ \ \ \ \ \ \ \ \ \ \textrm{(a)} \nonumber \\
P\mathbb{M}^{\mu\nu}P^{-1} & = &
{\mathcal{P}^{\mu}}_{\alpha}{\mathcal{P}^{\nu}}_{\beta}\mathbb{M}^{\alpha\beta},
\ \ \ \ \ \ \ \ \textrm{(b)}\\
T\mathbb{P}^{\mu}T^{-1} & =
& -{\mathcal{T}^{\mu}}_{\nu}\mathbb{P}^{\nu},\ \ \ \ \ \ \ \ \ \ \ \ \ \  \textrm{(c)} \nonumber \\
T\mathbb{M}^{\mu\nu}T^{-1} & = &
-{\mathcal{T}^{\mu}}_{\alpha}{\mathcal{T}^{\nu}}_{\beta}\mathbb{M}^{\alpha\beta},\
\ \ \ \ \ \textrm{(d)}\nonumber
\end{eqnarray}

Para ver como transforman los $\mathbb{J}^{\mu}$ notamos que bajo
transformaciones de Lorentz la densidad de Levi-Civita permanece
invariante. As\'i en concordancia con el hecho de que
$\mathbb{M}^{\mu\nu}$ es un tensor antisim\'etrico
$\mathbb{J}^{\alpha}$ es una densidad vetorial de peso 1; y
entonces

\begin{eqnarray}
P\mathbb{J}^{\mu}P^{-1} & = &
det(\mathcal{P}){\mathcal{P}^{\mu}}_{\nu}\mathbb{J}^{\nu},
\ \ \ \ \ \ \ \textrm{(a)} \nonumber \\
T\mathbb{J}^{\mu}T^{-1} & = &
-det(\mathcal{T}){\mathcal{T}^{\mu}}_{\nu}\mathbb{J}^{\nu}.\ \ \ \
\ \textrm{(b)}
\end{eqnarray}
Observamos, asi, que bajo paridad e inversi\'on temporal sucede
que

\begin{eqnarray}\label{eq:ap121}
P\mathbb{J}^{0}P^{-1} = -\mathbb{J}^{0}&\qquad , \qquad&T\mathbb{J}^{0}T^{-1} = -\mathbb{J}^{0},\ \ \   \textrm{(a)} \nonumber \\
P\mathbb{P}^{0}P^{-1} = \mathbb{P}^{0}&\qquad ,
\qquad&T\mathbb{P}^{0}T^{-1} = \mathbb{P}^{0}.\ \ \ \ \
\textrm{(b)}\label{j0transf}
\end{eqnarray}
Como $\mathbb{J}^{0}$ tiene que ver con las rotaciones alrededor
de un eje imaginario perpendicular al plano $x^1x^2$ con el
sentido de rotaci\'on positivo seg\'un $\hat{x}_1\times\hat{x}_2$,
observamos que al hacer la reflexi\'on $x^1 \rightarrow -x^1$ este
nuevo observador tiene el eje de rotaci\'on positivo en el sentido
opuesto. En otra direcci\'on al hacer la inversi\'on temporal el
nuevo observador ve a los objetos rotar en sentido contrario.
Estas dos afirmaciones estan expl\'icitas en (\ref{j0transf}a).
Las transformaciones de $\mathbb{P}^{0}$ en (\ref{j0transf}b) son
las que tienen sentido f\'isico tal como argumentamos
anteriormente.

Finalmente, el objeto $\mathbb{P}_{\mu}\mathbb{J}^{\mu}$ se
comporta como un pseudo escalar
\begin{eqnarray}
T\mathbb{P}_{\mu}\mathbb{J}^{\mu}T^{-1} & = &
det(\mathcal{T})\mathbb{P}_{\mu}\mathbb{J}^{\mu},\\
P\mathbb{P}_{\mu}\mathbb{J}^{\mu}P^{-1} & = &
det(\mathcal{P})\mathbb{P}_{\mu}\mathbb{J}^{\mu}.
\end{eqnarray}
\'Este tendr\'a importancia en la secci\'on siguiente.

\section{Casimires}

Cuando estudiamos representaciones del grupo de Poincar\'e, nos
interesa los casimires del grupo, estos resultan ser naturalmente
$\mathbb{P}_{\mu}\mathbb{P}^{\mu}$ y
$\mathbb{P}_{\mu}\mathbb{J}^{\mu}$. Para una part\'icula de masa m
en reposo estos tienen el valor~\cite{jackiwnair}\cite{binegar}

\begin{equation}
\mathbb{P}_{\mu}\mathbb{P}^{\mu}=-m^{2}\quad, \quad
\mathbb{P}_{\mu}\mathbb{J}^{\mu}=-m\mathbb{J}^{0}\equiv +ms,
\end{equation}
donde diremos que $s$ es el spin de la representaci\'on. Dado su
analog\'ia con el concepto de helicidad, en algunos casos se le
suele llamar ``helicidad'' de la representaci\'on, quedando claro
que no corresponde a una exitaci\'on sin masa como sucede en 3+1
dimensiones. Finalmente notamos que, teniendo en cuenta
(\ref{eq:ap121}), al hacer una transformaci\'on discreta de
paridad cambia el valor de $\mathbb{J}^{0}$ de $s$ a $-s$. Se
afirma, entonces, que una teor\'ia que describa una sola
exitaci\'on de masa $m$ y spin $s$ necesariamente ser\'a sensible
a las transformaciones impropias. Estas representaciones son
unidimensionales y satisfacen las condiciones de la capa de masa

$$(\mathbb{P}_{\mu}\mathbb{P}^{\mu} - m^{2})\vert \psi \rangle=0 \ \ \ \ \ \textrm{(a)},$$
y la de ``Pauli-Lubanski''

\begin{equation}
(\mathbb{P}_{\mu}\mathbb{J}^{\mu} - ms)\vert\psi \rangle=0 \ \ \ \
\ \textrm{(b)}.
\end{equation}

Cuando intentamos realizar alguna de estas representaciones, introducimos objetos
que transformen linealmente bajo el grupo de Lorentz homog\'eneo y entonces los grados
de libertad adicionales se eliminan dando condiciones suplementarias adicionales.

Recalcamos, para concluir, que en 2+1 dimensiones una part\'icula
de spin $s$ y masa $m$ tiene un s\'olo grado de
libertad~\cite{jackiwnair} (a diferencia del mismo caso en 3+1
dimensiones, donde tiene $2s+1$ grados de libertad).

\section{Realizaciones con s=1}

Para una realizaci\'on del grupo de Poincar\'e con s=1, usamos como objeto a un campo
vectorial $V^{\mu}$. \'Este transforma linealmente bajo el grupo de Lorentz.
Al operador $\mathbb{P}_{\mu}$ lo realizamos como

\begin{equation}
{(\mathbb{P}_{\mu})^{\alpha}}_{\beta}=i{\delta^{\alpha}}_{\beta}\partial_{\mu}
\end{equation}

y al operador $\mathbb{J}^{\mu}$ como

\begin{eqnarray}
{(\mathbb{J}_{\mu1})^{\alpha}}_{\beta} & = &
-{{\varepsilon^{\mu}}_{\lambda}}^{\rho}x^{\lambda}{(\mathbb{P}_{\rho})^{\alpha}}_{\beta}
+ i{\varepsilon^{\alpha\mu}}_{\beta}\ \ \ \ \  \textrm{(a)}, \nonumber \\
& \equiv & (L^{\mu})^{\alpha}_{\beta} +
(j^{\mu})^{\alpha}_{\beta}\ \ \ \  \ \ \ \ \ \ \ \  \textrm{(b)}
\end{eqnarray}
donde $(L^{\mu})^{\alpha}_{\beta}$ es la parte orbital usual.
$\mathbb{P}_{\mu}$ y $\mathbb{J}_{\mu 1}$ satisfacen, al actuar
sobre vectores, el \'algebra (\ref{eq:ap113})

\begin{eqnarray}\label{eq:ap125}
i[\mathbb{P}^{\mu},\mathbb{P}^{\nu}]^{\alpha}_{\beta} & =
& 0 \ \  \ \ \ \ \ \ \ \ \ \ \ \ \ \ \ \ \   \textrm{(a)}, \nonumber \\
i[\mathbb{P}^{\mu},\mathbb{J}^{\nu}_{1}]^{\alpha}_{\beta} & =
& -\varepsilon^{\mu\nu\lambda}(\mathbb{P}_{\lambda})^{\alpha}_{\beta}\ \ \ \ \ \ \textrm{(b)},\\
i[\mathbb{J}^{\mu}_{1},\mathbb{J}^{\nu}_{1}]^{\alpha}_{\beta} & =
&
-\varepsilon^{\mu\nu\lambda}(\mathbb{J}_{1\lambda})^{\alpha}_{\beta}\
\ \ \ \ \textrm{(c)}.\nonumber
\end{eqnarray}
La condici\'on de Pauli-Lubanski se ``lee'' ahora como

\begin{equation}
[(\mathbb{P}_{\mu}\mathbb{J}^{\mu}_{1}) -
ms\mathbb{I}]^{\alpha}_{\beta}V^{\beta}=0,
\end{equation}
(con $s=\pm 1$). En componentes

\begin{equation}
(-{\varepsilon^{\alpha\mu}}_{\beta}\partial_{\mu} -
ms{\delta^{\alpha}}_{\beta})V^{\beta}=0,
\end{equation}
de donde surgen las condiciones suplementarias
$\partial_{\mu}V^{\mu}=0$ y el v\'inculo \\
$V^{0}=-\frac{1}{ms}\varepsilon^{ij}\partial_{i}V_{j}$
dej\'andonos con un solo grado de libertad. Teniendo en cuenta la
transversalidad de $V^{\mu}$ obtendremos que

\begin{equation}\label{eq:ap128}
[-\mathbb{P}_{\mu}\mathbb{J}^{\mu}_{1} -
ms\mathbb{I}][\mathbb{P}_{\nu}\mathbb{J}^{\mu}_{1} -
ms\mathbb{I}]^{\alpha}_{\beta}V^{\beta}=0,
\end{equation}
que nos lleva a la condici\'on de la capa de masa

\begin{equation}
(-\Box + m^{2})V^{\alpha}=0
\end{equation}
tanto para $s=+1$ \'o $-1$.

\section{Realizaciones con $s=2$}

Para $s=2$, el objeto que usaremos es un tensor $h_{\mu\nu}$
sim\'etrico. El operador
${(\mathbb{J}_{2}^{\mu})_{\rho\sigma}}^{\alpha\beta}$ es ahora

\begin{eqnarray}
{(\mathbb{J}_{2}^{\mu})_{\alpha\beta}}^{\gamma\sigma} & = &
-i{{\varepsilon^{\mu}}_{\lambda}}^{\rho}x^{\lambda}
{(\mathbb{I}_{s})_{\alpha\beta}}^{\gamma\sigma}\partial_{\mu}+
\nonumber\\
& & +\frac{i}{2}(\delta^{\gamma}_{\alpha}{\varepsilon_{\beta}}^{\mu\sigma} +
\delta^{\gamma}_{\beta}{\varepsilon_{\alpha}}^{\mu\sigma} +
+\delta^{\sigma}_{\alpha}{\varepsilon_{\beta}}^{\mu\gamma} +
\delta^{\sigma}_{\beta}{\varepsilon_{\alpha}}^{\mu\gamma}), \\
& \equiv & -{{\varepsilon^{\mu}}_{\lambda}}^{\rho}x^{\lambda}
{(\mathbb{P}_{\mu})_{\alpha\beta}}^{\gamma\sigma} +
{(j^{\mu})_{\alpha\beta}}^{\gamma\sigma},\label{j2}
\end{eqnarray}
donde ${{(\mathbb{I}_{s})}_{\alpha\beta}}^{\gamma\sigma}=
\frac{1}{2}(\delta_{\alpha}^{\gamma}\delta_{\beta}^{\sigma} +
\delta_{\alpha}^{\sigma}\delta^{\gamma}_{\beta})$ y el primer
t\'ermino es la parte orbital. Es inmediato comprobar que actuando
sobre tensores sim\'etricos de segundo orden estos operadores
satisfacen el \'algebra del grupo de Poincar\'e.

\begin{eqnarray}
i{[\mathbb{P}^{\mu},\mathbb{P}^{\nu}]_{\alpha\beta}}^{\rho\sigma} & = & 0
 \ \ \ \ \ \ \ \ \ \ \ \ \ \ \ \ \ \ \ \ \ \textrm{(a)}, \nonumber \\
i{[\mathbb{P}^{\mu},\mathbb{J}^{\nu}_{2}]_{\alpha\beta}}^{\rho\sigma} & =
& -\varepsilon^{\mu\nu\lambda}{(\mathbb{P}_{\lambda})_{\alpha\beta}}^{\rho\sigma}
\ \ \ \ \ \ \textrm{(b)},\\
i{[\mathbb{J}^{\mu}_{2},\mathbb{J}^{\nu}_{2}]_{\alpha\beta}}^{\rho\sigma}
& = &
-\varepsilon^{\mu\nu\lambda}{(\mathbb{J}_{2\lambda})_{\alpha\beta}}^{\rho\sigma}\
\ \ \ \ \textrm{(c)},\nonumber
\end{eqnarray}

La condici\'on de Pauli-Lubanski es ahora

\begin{equation}\label{eq:ap131}
{[(\mathbb{P}_{\mu}\mathbb{J}^{\mu}_{2}) -
ms\mathbb{I}]_{\alpha\beta}}^{\rho\sigma}h_{\rho\sigma}=0,
\end{equation}
con $s=\pm2$.

Las condiciones suplementarias que impondremos son, adem\'as de la
simetr\'ia, que el tensor sea transverso y sin traza

\begin{equation}
\partial^{\mu}h_{\mu\nu}=0 \quad , \quad
\eta^{\mu\nu}h_{\mu\nu}=0.
\end{equation}
As\'i, la ecuaci\'on (\ref{eq:ap131}) en componentes es

\begin{equation}
-2{\varepsilon_{\alpha}}^{\mu\sigma}\partial_{\mu}h_{\sigma\beta}^{Tt}
- msh_{\alpha\beta}^{Tt}=0, \label{eq. s2}
\end{equation}
donde $h_{\alpha\beta}^{Tt}$ denota a un tensor sim\'etrico,
transverso y sin traza. \'Este tiene dos componentes
independientes y puede verse que de \'estas s\'olo se propaga una
de ellas en (\ref{eq. s2}).

Finalmente la condici\'on de la capa de masa se obtiene si procedemos al igual
que en (\ref{eq:ap128}). En componentes

\begin{equation}
(-2{\varepsilon_{\alpha}}^{\mu\theta_{1}}\delta_{\beta}^{\theta_{2}}\partial_{\mu}
- ms\delta_{\alpha}^{\theta_{1}}\delta_{\beta}^{\theta_{2}})
(2{\varepsilon_{\theta_{1}}}^{\nu\rho}
\delta_{\theta_{2}}^{\sigma}\partial_{\nu} -
ms\delta_{\theta_{1}}^{\rho}\delta_{\theta_{2}}^{\sigma})h_{\rho\sigma}^{Tt}
=4(-\Box
+ m^{2})h_{\rho\beta}^{Tt}
\end{equation}
para $s=\pm2$.

\section{Realizaciones con $s=n$, con $n$ entero}
Los resultados de la secci\'on anterior pueden generalizarse para
$s=n$ entero . El objeto fundamental ser\'a un tensor de orden n
sim\'etrico $h_{\mu_1\mu_2\cdots\mu_n}$, al cual imponemos que sea
transverso y sin traza en un par de \'indices
\begin{equation}
\partial^{\mu}h_{\mu\mu_1\mu_2\cdots\mu_{n-1}}=0, \quad
\eta^{\mu\nu}h_{\mu\nu\mu_1\mu_2\cdots\mu_{n-2}}=0.
\end{equation}
Como un tensor sim\'etrico de orden $n$ tiene
$\frac{(n+1)(n+2)}{2}$ se obtiene, teniendo en cuenta las
condiciones subsidiarias, que el n\'umero de componentes
independientes es
$$\frac{(n+1)(n+2)}{2}-\frac{n(n+1)}{2}-\frac{(n-1)n}{2}=2$$.

El operador asociado al momento angular
${(\mathbb{J}_{n}^{\mu})_{\sigma_1\sigma_2\cdots\sigma_n}}^{\rho_1\rho_2\cdots\rho_n}$
se generaliza como

\begin{eqnarray}
{(\mathbb{J}_{n}^{\mu})_{\sigma_1\sigma_2\cdots\sigma_n}}^{\rho_1\rho_2\cdots\rho_n}&=&
-i{{\varepsilon^{\mu}}_{\lambda}}^{\nu}x^{\lambda}
{(\mathbb{I}_{s})_{\sigma_1\sigma_2\cdots\sigma_n}}
^{\rho_1\rho_2\cdots\rho_n}\partial_{\nu}+\cdots
\nonumber\\
& &
+\frac{i}{n!(n-1)!}{\varepsilon_{(\sigma_1}}^{\mu(\rho_1}\delta^{\rho_2}_{\sigma_2}\cdots
\delta^{\rho_n)}_{\sigma_n)}, \nonumber\\
& \equiv & -{{\varepsilon^{\mu}}_{\lambda}}^{\nu}x^{\lambda}
{(\mathbb{P}_{\nu})_{\sigma_1\sigma_2\cdots\sigma_n}}^{\rho_1\rho_2\cdots\rho_n}
+
{(j^{\mu})_{\sigma_1\sigma_2\cdots\sigma_n}}^{\rho_1\rho_2\cdots\rho_n},\label{sn}
\end{eqnarray}
con
${{(\mathbb{I}_{s})}_{\sigma_1\sigma_2\cdots\sigma_n}}^{\rho_1\rho_2\cdots\rho_n}=
\frac{1}{n!}\delta_{(\sigma_1}^{\rho_1}\delta_{\sigma_2}^{\rho_2}
\cdots\delta_{\sigma_n)}^{\rho_n}$, donde
$(\mu_1\mu_2\cdots\mu_n)$ significa simetrizaci\'on en los $n$
\'indices. El primer t\'ermino en (\ref{sn}) es la parte orbital,
en analog\'ia con los casos vistos. Se comprueba que actuando
sobre tensores sim\'etricos de orden n estos operadores satisfacen
el \'algebra del grupo de Poincar\'e.

Lo que sigue es la genaralizaci\'on de la condici\'on de
Pauli-Lubanski que es ahora

\begin{equation}
{{[(\mathbb{P}_{\mu}\mathbb{J}^{\mu}_{n}) -
ms\mathbb{I}]}_{\sigma_1\sigma_2\cdots\sigma_n}}
^{\rho_1\rho_2\cdots\rho_n}h_{\rho_1\rho_2\cdots\rho_n}=0,
\end{equation}
con $s=\pm n$. Teniendo en cuenta las propiedades y condiciones
subsidiarias de $h_{\rho_1\rho_2\cdots\rho_n}$, \'esta ecuaci\'on
queda en componentes como
\begin{equation}
-n{\varepsilon_{\rho_1}}^{\mu\sigma}\partial_{\mu}h_{\sigma\rho_2\rho_3\cdots\rho_n}^{Tt}
- msh_{\rho_1\rho_2\cdots\rho_n}^{Tt}=0, \label{eq. sncomponentes}
\end{equation}

La condici\'on de la capa de masa
$(-\Box+m^2)h_{\rho_1\rho_2\cdots\rho_n}^{Tt}=0$ se obtiene
directamente de forma an\'aloga a los casos de spin 1 y 2. Para
ver que finalmente se propaga una sola excitaci\'on notamos que
$h_{\rho_1\rho_2\cdots\rho_n}^{Tt}$ se descompone en dos partes
irreducibles
\begin{equation}
h_{\rho_1\rho_2\cdots\rho_n}^{Tt(\pm)}=\frac{1}{2}\left(h_{\rho_1\rho_2\cdots\rho_n}^{Tt}
\mp {\varepsilon_{\rho_1}}^{\mu\sigma}{\frac{\partial_{\mu}}
{\Box^{\frac{1}{2}}}}
h_{\sigma\rho_2\rho_3\cdots\rho_n}^{Tt}\right),
\end{equation}
con lo que (\ref{eq. sncomponentes}) queda como ($s=n$)
\begin{equation}
(\Box^\frac{1}{2}-m)h_{\rho_1\rho_2\cdots\rho_n}^{Tt(+)}
+(\Box^\frac{1}{2}+m)h_{\rho_1\rho_2\cdots\rho_n}^{Tt(-)}=0,
\end{equation}
de donde observamos que en la capa de masa solo una de las partes
se propaga.

Para finalizar este cap\'itulo se\~nalamos que, aunque en este
trabajo nos abocaremos a teor\'ias de spines $1$ y $2$, hay
estudios realizados con distintos modelos que constituyen
realizaciones para $s=3,4,\cdots$~\cite{adel}\cite{vasiliev}.

\chapter{Dualidad para teor\'{\i}as con spin 1}

En este cap\'itulo mostraremos que dos realizaciones posibles con
$s=1$, estan conectadas por una transformaci\'{o}n de dualidad.
Los modelos difieren en las simetr\'{\i}as de calibre que tiene
cada uno, pudiendo conectarlos igualmente fijando el calibre en
uno de los modelos. El proceso de dualidad que seguiremos y que
luego generalizaremos a spin $2$ est\'a desarrollado en la
referencia~\cite{jorge}.

A nivel de las funciones de partici\'{o}n se ver\'{a} que
\'{e}stas difieren en un factor topol\'{o}gico el cual es sensible
a la topolog\'{\i}a de la variedad base.

En la discusi\'{o}n tambi\'en veremos que, as\'{\i} como se sabe
que las acciones de los modelos, que describen part\'{\i}culas
masivas con un spin dado, deben ser sensibles a paridad e
inversi\'on temporal, puede mostrarse que el modelo de Proca se
desacopla en dos modelos que corresponden a exitaciones masivas
con spines opuestos, cuya presencia es necesaria para tener un
modelo invariante bajo estas transformaciones discretas.

Este cap\'{\i}tulo servir\'{a} de base para la discusi\'{o}n del cap\'{\i}tulo
siguiente donde trataremos la dualidad entre modelos de spin 2.

\section{Los modelos como realizaciones del \'{a}lgebra de Poincar\'{e}}

Para la descripci\'{o}n de una teor\'{\i}a masiva con spin 1,
tomamos como objeto a un campo vectorial $V^{\mu}$. Tal como
discutimos en el cap\'itulo anterior a condici\'{o}n de
Pauli-Lubanski se escribe en componentes como

\begin{equation} \label{eq:ec2}
(-\varepsilon^{\alpha\mu}_{\beta} \partial_{\mu} -
ms\delta^{\alpha}_{\beta})V^{\beta}=0,
\end{equation}
de donde ocurre que $\partial_{\mu}V^{\mu}=0$. La teor\'{\i}a que
tiene a (\ref{eq:ec2}) como ecuaci\'{o}n de movimiento corresponde
al modelo actual (SD)~\cite{PK}\cite{DJself}

\begin{equation}
S_{SD}= - \frac{ms}{2} \int d^{3}x (V_{\mu}
\varepsilon^{\mu\nu\lambda} \partial_{\nu} V_{\lambda} + ms
V_{\mu}V^{\mu}),
\end{equation}
con $s=\pm 1$. Puede verse que los generadores asociados a la
invariancia de la teor\'{\i}a por el grupo $ISO(2,1)$  satisfacen
el \'{a}lgebra de \'{e}ste grupo y que el spin es efectivamente $s
(+1 \ \acute{o} -1)$~\cite{girotti}\cite{piomonica}.

En $2+1$ dimensiones, un vector es el dual de un tensor
antisim\'{e}trico, digamos $F_{\mu\nu}$. As\'{\i} la ecuaci\'{o}n
(\ref{eq:ec2}) se escribe igualmente como

\begin{equation}\label{eq:ec4}
(\varepsilon^{\alpha\mu}_{\beta} \partial_{\mu} - ms
\delta^{\alpha}_{\beta})(\frac{1}{2}\varepsilon^{\beta\rho\sigma}
F_{\rho\sigma})=0.
\end{equation}
La condici\'{o}n $\partial_{\mu}V^{\mu}=0$ es ahora

\begin{equation}\label{eq:ec5}
\varepsilon^{\mu\nu\lambda} \partial_{\mu} F_{\nu\lambda}=0.
\end{equation}
cuya soluci\'{o}n local es $F_{\mu \nu}(a)= \partial_{\mu} a_{\nu}
- \partial_{\nu} a_{\mu}$. Resulta que (\ref{eq:ec5}) es
justamente la identidad de Bianchi que satisface el tensor de
Maxwell, $F_{\mu\nu}$, en dimensi\'{o}n 2+1.

As\'{\i} (\ref{eq:ec4}) y (\ref{eq:ec5}) se juntan en una sola ecuaci\'{o}n como

\begin{equation}\label{eq:ec6}
{(\varepsilon^{\alpha\mu}}_{\beta} \partial_{\mu} - ms
\delta^{\alpha}_{\beta})f^{\beta}(a)=0,
\end{equation}
con $f^{\mu}(a)=
\frac{1}{2}\varepsilon^{\mu\nu\lambda}F_{\nu\lambda}(a)$. Estas
son la ecuaciones de movimiento del modelo Topol\'{o}gico Masivo
(TM)~\cite{DJT} cuya acci\'{o}n es

\begin{eqnarray}\label{eq:ec7}
S_{TM} & = & \int \rd^{3}x (-\frac{1}{4} F_{ \mu \nu}(a) F^{ \mu
\nu}(a) + \frac{ms}{2} a_{\mu} \varepsilon ^{\mu\nu\lambda}
\partial_{\nu} a_{\lambda},
\ \textrm{(a)} \nonumber \\
& = & \frac{1}{2} \int \rd^{3}x (f_{\mu}(a)f^{\mu}(a) + ms
a_{\mu}f^{\mu}(a)), \  \ \ \ \textrm{(b)}
\end{eqnarray}
donde $s=\pm 1$. Es importante recalcar que si empezaramos con la
acci\'{o}n autodual  con $V^{\mu}=\frac{1}{2}\varepsilon^{\mu \nu
\lambda} F_{\nu \lambda}(a)$, las ecuaciones de movimiento seran
igualmente equivalentes a las del modelo TM a nivel local.

Las ecuaciones (\ref{eq:ec7}) son invariantes bajo las
transformaciones de calibre $\delta
a_{\mu}=\partial_{\mu}\lambda$. La componente $0$ de
(\ref{eq:ec6}) es la ley de Gauss modificada, que luego en el
procedimiento Hamiltoniano pasa a ser el generador de las
transformaciones de calibre de los $a_{i}$, siendo $a_{0}$ el
multiplicador de Lagrange asociado. Una vez fijado el calibre nos
quedar\'{a} un s\'{o}lo grado de libertad, como debe de ser.

La teor\'{\i}a TM es tambi\'{e}n una teor\'{\i}a covariante bajo
el grupo $ISO(2,1)$ y tiene spin $ s (+1 \ \acute{o} -1
)$~\cite{DJT}.

Tal como vimos en el cap\'itulo anterior la condici\'{o}n de
Pauli-Lubanski es sensible a paridad e inversi\'on temporal. Tanto
el modelo SD como el TM poseen esta caracter\'{\i}stica si
consideramos transformaciones discretas P y T. Bajo estas
transformaciones sucede que

\begin{eqnarray}
P a_{0}(x)P^{-1} & = & a_{0}(x_{P}), \nonumber \\
P a_{1}(x)P^{-1} & = & -a_{1}(x_{P}),  \  \  \  \  \  \  \  \  \  \ \textrm{(a)} \nonumber\\
P a_{2}(x)P^{-1} & =& a_{2}(x_{P}),  \nonumber \\ \\ \nonumber
Ta_{0}(x)T^{-1} & = &  a_{0}(x_{T}), \nonumber \\
Ta_{i}(x)T^{-1} & = & - a_{i}(x_{T}), \  \  \  \  \  \  \  \  \  \
\ \textrm{(b)} \nonumber
\end{eqnarray}\nonumber
\nonumber
con \nonumber
\nonumber
\begin{eqnarray}
x^{\mu}_{P} & = & (x^{0},-x^{\prime},x^{2}), \nonumber \\
x^{\mu}_{T} & = & (-x^{0},x^{i}),  \  \  \  \  \  \  \  \  \  \
\  \  \ \ \ \  \ \ \  \ \textrm{(c)} \nonumber
\end{eqnarray}
igualmente sucede con $V_{\mu}(x)$ en el modelo autodual. En ambos
casos el t\'{e}rmino de Cheen-Simons (CS) $a_{\mu}\varepsilon^{\mu
\nu \lambda} \partial_{\nu}a_{\lambda}$ (\'{o}
$V_{\mu}\varepsilon^{\mu \nu \lambda}\partial_{\nu}V_{\lambda}$)
cambia de signo. As\'{\i} una transformaci\'{o}n combinada PT deja
invariante a \'este t\'{e}rmino. Tenemos que los modelos SD y, TM
son sensibles a paridad e inversi\'{o}n temporal. El spin de las
exitaciones en el modelo SD y TM es $s=\frac{m}{\vert m
\vert}$~\cite{DJT}\cite{PJA}\cite{girotti}\cite{piomonica},
as\'{\i} al cambiar del signo del t\'{e}rmino de CS sencillamente
cambia el spin de la exitaci\'{o}n, por tanto las transformaciones
impropias nos llevan a un modelo de igual tipo pero con spin
opuesto.

Una diferencia notoria de caracter topol\'ogico entre los modelos
$TM$ y $SD$ es que en la ecuaci\'on (\ref{eq:ec2}) del modelo $SD$
las \'unicas soluciones correspondientes a $F_{\mu\nu}(V)=0$ son
las triviales ($V_{\mu}=0$), en cambio para la $TM$ en
(\ref{eq:ec6}) las soluciones triviales y no triviales estan
expl\'icitamente inclu\'idas. Tenemos, por tanto, que el espacio
de soluciones de ambos modelos son diferentes y difieren al menos
en un sector asociado con las soluciones no triviales de
$F_{\mu\nu}=0$ el cual est\'a \'intimamente relacionado con el
primer grupo de cohomolog\'ia de la variedad base. \'Este espacio
es precisamente el espacio de soluciones del modelo de
Chern-Simons puro~\cite{pioalvaro,sen}.

Si queremos una teor\'{\i}a que preserve P y T debemos tener
presente dos exitaciones con spines opuestos (e igual masa).
\'{E}ste es el caso del modelo de Proca

\begin{equation} \label{eq:ec9}
S_{P}=\int \rd^{3}x (-\frac{1}{4} F_{\mu\nu} F^{\mu\nu} -
\frac{m^{2}}{2}a_{\mu}a^{\mu}),
\end{equation}
el cual puede verse que es invariante bajo estas transformaciones
discretas. Para corroborar la afirmaci\'{o}n hecha, reescribimos
la acci\'{o}n (\ref{eq:ec9}) a primer orden

\begin{equation} \label{eq:ec10}
\bar{S_{P}}= \int \rd^{3}x (\varepsilon^{\mu \nu
\lambda}b_{\mu}\partial_{\nu}a_{\lambda} - \frac{1}{2}b_{\mu}
b^{\mu} - \frac{m^{2}}{2}a_{\mu}a^{\mu}),
\end{equation}
donde $b_{\mu}$ es un pseudo-vector para mantener la invariancia
bajo P y T en la acci\'{o}n (\ref{eq:ec10}). Haciendo variaciones
respecto a $b^{\mu}$ obtenemos $b^{\mu}=f^{\mu}(a)$ que al
sustituir en $\bar{S_{P}}$ nos lleva a  (\ref{eq:ec9}). Si hacemos
el cambio

\begin{eqnarray}
a_{\mu} & = & \frac{1}{\sqrt{2}} (a_{1\mu} + a_{2\mu}), \ \ \textrm{(a)} \nonumber \\
b_{\mu} & = & \frac{m}{\sqrt{2}} (a_{1\mu} - a_{2\mu}),  \ \
\textrm{(b)}
\end{eqnarray}
y sustituimos en (\ref{eq:ec10}) obtendremos, eliminando las
contribuciones de borde

\begin{eqnarray} \label{eq:ec12}
\bar{S}_{P}\Big|_{10}  =  -\frac{1}{2}\int d^{3}x \lbrack -\varepsilon^{\mu \nu \lambda}a_{1\mu}\partial_{\nu}a_{1\lambda} + ma_{1\mu}a^{\mu}_{1} + \nonumber \\
 + \varepsilon^{\mu \nu \lambda} a_{2\mu}\partial_{\nu}a_{2\lambda} +
 ma_{2\mu}a^{\mu}_{2}\rbrack,
\end{eqnarray}
que corresponde a dos modelos SD desacoplados. Uno con $s=-1$ (el
de $a_{1\mu}$) y otro con $s=+1$ (el de $a_{2\mu}$).

Las transformaciones P y T las definimos ahora de forma que
incluyan un intercambio de los campos $a_{1\mu}$ y $a_{2\mu}$.
As\'{\i} el car\'{a}cter pseudo-tensorial de $b_{\mu}$ aparece y
(\ref{eq:ec12}) queda invariante. En el cap\'itulo siguiente
cuando veamos las teor\'{\i}as con $s=2$ veremos que esta
situaci\'{o}n reaparece con el modelo de Einstein-Fierz-Pauli.

Pasamos ahora a conectar los modelos SD y TM por dualidad.

\section{Dualidad entre los modelos SD y TM}

El esquema de dualidad que utilizaremos es el
siguiente~\cite{jorge}~\cite{busher}: partiendo de la teor\'{\i}a
SD con un campo vectorial $a_{\mu}$, introducimos un acoplamiento
de la teor\'{\i}a original con otro campo vectorial $A_{\mu}$ el
cual forzamos a que corresponda a una conexi\'{o}n plana
$(F_{\mu\nu}(A)=0)$ por la v\'{\i}a de un multiplicador de
Lagrange $B_{\mu}$. La teor\'{\i}a acoplada con $a_{\mu}$,
$A_{\mu}$ y $B_{\mu}$ es equivalente localmente al modelo SD. . La
teor\'{\i}a dual corresponder\'{a} a la teor\'{\i}a resultante
luego de eliminar a $a_{\mu}$ y  $A_{\mu}$ usando las ecuaciones
de movimiento. Este proceso de eliminaci\'{o}n es equivalente al
que hici\'{e}ramos en la integral funcional.

Empezamos, as\'{\i} con el modelo autodual

\begin{equation}
S_{SD}= -\frac{m}{2} \int \rd^{3}x \lbrack \varepsilon^{\mu \nu
\lambda}a_{\mu}\partial_{\nu}a_{\lambda} + ma_{\mu}a^{\mu}\rbrack.
\end{equation}
En variedades triviales el primer t\'ermino de esta acci\'on es
equivalente a otro escrito de la forma $\varepsilon^{\mu \nu
\lambda}(a_{\mu}+A_{\mu})\partial_{\nu}(a_{\lambda}+A_{\lambda})$
si el campo auxiliar $A_{\mu}$ le es impuesta la condici\'on
$F_{\mu\nu}(A)=0$. En variedades no triviales no tiene porque ser
as\'i y por esta v\'ia a las propiedades topol\'ogicas no
triviales pueden ser introducidas dentro de la formulaci\'on.

El modelo acoplado ser\'{a}

\begin{eqnarray}\label{eq:ec14}
\tilde{S}_{SD}= -\frac{m}{2} \int d^{3}x
\lbrack \varepsilon^{\mu \nu \lambda}(a_{\mu} + A_{\mu})\partial_{\nu}(a_{\lambda} +
A_{\lambda})+\\
 +ma_{\mu}a^{\mu} + 2B_{\mu}\varepsilon^{\mu \nu \lambda}\partial_{\nu}A_{\lambda}
 \rbrack, \nonumber
\end{eqnarray}
donde el campo $B_{\mu}$ aparece como un multiplicador que forza a
$A_{\mu}$ a ser una conexi\'{o}n plana ($\varepsilon^{\mu \nu
\lambda}\partial_{\lambda}A_{\lambda}=0$).

Las ecuaciones de movimiento del modelo $\tilde{S}_{SD}$, al hacer
variaciones independientes en los campos, son:

\begin{eqnarray}\label{eq:ec15}
-m^{2}a^{\mu} - m\varepsilon^{\mu \nu \lambda}\partial_{\nu}(a_{\lambda} + A_{\lambda}) & = & 0 \ \ \textrm{(a)} \nonumber \\
\varepsilon^{\mu \nu \lambda}\partial_{\nu}(a_{\lambda} + A_{\lambda} + B_{\lambda}) & = & 0 \ \ \textrm{(b)} \\
\varepsilon^{\mu \nu \lambda}\partial_{\nu}A_{\lambda} & = & 0 \ \ \textrm{(c)} \nonumber
\end{eqnarray}
las cuales son invariantes bajo las transformaciones de calibre

\begin{equation}
\delta A_{\mu}=\partial_{\mu} \Lambda_{1} \qquad \delta
B_{\mu}=\partial_{\mu} \Lambda_{2}.
\end{equation}
La invariancia de calibre asociado a $A_{\mu}$ y la ecuaci\'on
(\ref{eq:ec15}c) permiten, localmente, eliminarlo de la acci\'on
($A_{\mu}=\partial_{\mu}\lambda \sim 0$) recuperando la acci\'on
autodual usual (la ecuaci\'on \ref{eq:ec15}b permitir\'a, ahora
identificar $a_{\mu}$ con $B_{\mu}$). Nos interesa ver cual es el
modelo asociado con s\'olo el campo $B_{\mu}$. Llamemos

\begin{equation}
\tilde{A}_{\mu}\equiv a_{\mu} + A_{\mu} + B_{\mu},
\end{equation}
la ecuaci\'on (\ref{eq:ec15}b) dice que

\begin{equation}\label{eq:ec18}
\varepsilon^{\mu \nu \lambda}\partial_{\nu}\tilde{A}_{\lambda}=0,
\end{equation}
y escogemos el calibre

\begin{equation}
\partial_{\mu}\tilde{A}^{\mu}=0.
\end{equation}

De (\ref{eq:ec18}) tenemos que $\tilde{A}_{\mu}=0$, as\'i

\begin{equation} \label{eq:ec19}
B_{\mu}=-(a_{\mu} + A_{\mu}).
\end{equation}
Sustituyendo (\ref{eq:ec18}) y  (\ref{eq:ec19}) en (\ref{eq:ec14})
llegamos a

\begin{equation}\label{eq:ec20}
\tilde{S}_{SD}\Big|_{\tilde{A}_{\mu}=0} = -\frac{m}{2}\int
\rd^{3}x (-\varepsilon^{\mu \nu \lambda}
B_{\mu}\partial_{\nu}B_{\lambda} + ma_{\mu}a^{\mu} -
2B_{\mu}\varepsilon^{\mu \nu \lambda}\partial_{\nu}a_{\lambda}),
\end{equation}
ahora el papel de $a_{\mu}$ es como un campo auxiliar. El modelo
resultante resulta ser una versi\'on a primer orden de la teor\'ie
TM. De hecho, las ecuaciones de movimiento de $a_{\mu}$ son

$$-m^{2}a^{\mu} + m\varepsilon^{\mu\nu\lambda}\partial_{\nu}B_{\lambda} = 0$$

\'o
\begin{equation}\label{eq:ec21}
a^{\mu} =\frac{1}{m}\varepsilon^{\mu \nu
\lambda}\partial_{\nu}B_{\lambda},
\end{equation}
que al sustituir en (\ref{eq:ec20}) nos llevan al modelo TM en
funci\'on de $B_{\mu}$

\begin{eqnarray}\label{eq:ec22}
\Big(\tilde{S}_{SD}\Big|_{\tilde{A}_{\mu}=0}\Big) \Big|_{21} & \equiv & S_{TM} \nonumber \\
& = & \int \rd^{3}x (-\frac{1}{4} F_{\mu \nu}(B)F^{\mu \nu}(B) +
\frac{m}{2}\varepsilon^{\mu \nu
\lambda}B_{\mu}\partial_{\nu}B_{\lambda}),
\end{eqnarray}

Los c\'alculos ''on shell'' hechos hasta ahora pueden hacerse
desde el punto de vista de la integraci\'on funcional y nos
permitir\'an obtener una relaci\'on entre las funciones de
partici\'on de los modelos SD y TM. Notemos que (\ref{eq:ec14})
puede escribirse como:

\begin{eqnarray}
\tilde{S}_{SD}&=&-\frac{m}{2} \int \rd^{3}x [\varepsilon^{\mu \nu
\lambda}
a_{\mu}\partial_{\nu}a_{\lambda} + ma_{\mu}a^{\mu} + \nonumber \\
 &&\qquad +(2a_{\mu} + 2B_{\mu} + A_{\mu})\varepsilon^{\mu \nu
 \lambda}\partial_{\nu}A_{\lambda}],\label{SZ1}\\
 &=& -\frac{m}{2}\int
\rd^{3}x [-\varepsilon^{\mu \nu \lambda}
B_{\mu}\partial_{\nu}B_{\lambda} + ma_{\mu}a^{\mu} -
2B_{\mu}\varepsilon^{\mu \nu
\lambda}\partial_{\nu}a_{\lambda}+\nonumber \\
&&\qquad +(a_{\mu} + B_{\mu} + A_{\mu})\varepsilon^{\mu \nu
 \lambda}\partial_{\nu}(a_{\mu} + B_{\mu} + A_{\mu})],
 \label{SZ2}
\end{eqnarray}
donde observamos que si hacemos el cambio de variable
$2\tilde{A}_{\mu} - A_{\mu}= \tilde{B}_{\mu}$ en (\ref{SZ1}) nos
queda el modelo SD mas un t\'ermino BF. La funci\'on de
partici\'on de este \'ultimo es la del modelo Cheen-Simons (CS)
puro al cuadrado~\cite{thompson}. As\'i

\begin{equation}
Z_{\tilde{S}_{SD}} \propto Z_{SD} \cdot (Z_{CS})^{2}
\end{equation}
la parte $Z_{CS}$ tiene car\'acter estrictamente topol\'ogico y
est\'a relacionado con la torsi\'on de Ray-Singer~\cite{thompson}.
Por otro en (\ref{SZ2}) observamos que en la integral funcional es
posible hacer la integraci\'on en $\tilde{A}_{\mu}$
(entendi\'endose que deba hacerse todo el proceso de fijaci\'on de
calibre) produci\'endose un factor $Z_{CS}$ y el resto es la
acci\'on $TM$. As\'i

\begin{equation}
Z_{\tilde{S}_{SD}} \propto Z_{TM} \cdot Z_{CS}.
\end{equation}
Tenemos, entonces, que
\begin{equation}
Z_{TM} \propto Z_{SD} \cdot Z_{CS}
\end{equation}
que es un resultado
conocido~\cite{pioalvaro}\cite{jorge}\cite{sen}.

Conclu\'imos que el modelo SD vectorial

$$S_{SD}=-\frac{m}{2} \int \rd^{3}x (\varepsilon^{\mu \nu \lambda}a_{\mu}\partial_{\nu}a_{\lambda} +ma_{\mu}a^{\mu})$$
es dual al modelo TM vectorial

$$ S_{TM}=\int \rd^{3}x (-\frac{1}{4}F_{\mu\nu}F^{\mu\nu} + \frac{m}{2}\varepsilon^{\mu \nu \lambda}a_{\mu}\partial_{\nu}a_{\lambda})$$
y sus funciones de partici\'on difieren en un factor topol\'ogico
asociado a la torsi\'on de Ray-Singer cuyo valor es sensible a las
propiedades topol\'ogicas de la variedad base. Cuando la
topolog\'ia es trivial estos modelos son completamente
equivalentes~\cite{DJself}~\cite{rita}~\cite{PJA}
~\cite{pioalvaro}~\cite{jorge}~\cite{sen}.

\chapter{Dualidad para teor\'{\i}as de spin 2}

En este cap\'itulo se presenta el resultado crucial de este
trabajo, donde se generaliza la dualidad existente entre los
modelos de spin $1$ a los de spin $2$, reafirmando el hecho ya
conocido de que los fen\'omenos que pueden predecirse para las
teor\'ias vectoriales tienen su an\'alogo en los modelos con spin
$2$~\cite{PJA}.

Veremos que los modelos asociados a la condicí\'on de
Pauli-Lubanski en este caso resultan ser tres: el modelo autodual
de spin 2 (SD)~\cite{ArKh}, el modelo intermedio (TI)~\cite{ArKh}
y el modelo topol\'ogico masivo linealizado~\cite{DJT}. Luego
pasaremos a la generalizaci\'on del m\'etodo usado en la
referencia \cite{jorge} a estos modelos. La notaci\'on que
utilizamos est\'a expuesta en el Ap\'endice.

\section{Las acciones involucradas y su relaci\'on con la condici\'on de Pauli-Lubanski}

En la discusi\'on que dimos relativa a las realizaciones del
\'algebra de Poincar\'e con $s=2$ el objeto considerado fu\'e un
tensor de segundo orden, sim\'etrico, transverso y sin traza
\begin{eqnarray}
h_{\mu\nu}^{Tt}= h_{\nu\mu}^{Tt},\qquad &
\partial^{\mu}h_{\mu\nu}^{Tt}=0,\qquad & {h^{\mu}}_{\mu}^{Tt}=0,
\end{eqnarray}
y la parte asociada al spin del momento angular, la realizamos con
el operador (ver (\ref{j2}))

\begin{equation}
{(\mathbb{J}^{\mu}_{2})_{\alpha\beta}}^{\gamma\sigma}=\frac{i}{2}(\delta_{\alpha}^{\gamma}{\varepsilon^{\sigma\mu}}_{\beta}
+ \delta^{\sigma}_{\alpha}{\varepsilon^{\alpha\mu}}_{\beta} +
\delta^{\gamma}_{\beta}{\varepsilon^{\sigma\mu}}_{\alpha} +
\delta^{\sigma}_{\beta}{\varepsilon^{\gamma\mu}}_{\alpha}).
\end{equation}
As\'i, la condici\'on de Pauli-Lubanski queda como

\begin{equation}\label{eq:cap22}
\frac{1}{2}{[(\mathbb{J}^{\mu}\mathbb{P}_{\mu}) \pm
2m\mathbb{I}]_{\alpha\beta}}^{\gamma\sigma}h_{\alpha\sigma}^{Tt}=0.
\end{equation}
\'Este objeto sim\'etrico, transverso y sin traza, tiene dos
componentes independientes. Sin embargo en (\ref{eq:cap22}) puede
verse que s\'olo uno de ellos se propaga. De hecho en este objeto
podemos distinguir las partes $h_{\alpha\beta}^{(+)Tt}$ y
$h_{\alpha\beta}^{(-)Tt}$~\cite{ArKh}\cite{PJA}

\begin{equation}\label{eq:cap23}
h_{\alpha\beta}^{(\pm) Tt}= \frac{1}{2}[h_{\alpha\beta}^{Tt}\pm
\frac{1}{2}({\xi_{\alpha}}^{\gamma}\delta_{\beta}^{\sigma} +
{\xi_{\beta}}^{\gamma}\delta_{\alpha}^{\sigma})h_{\gamma\sigma}^{Tt}],
\end{equation}
con
${\xi_{\alpha}}^{\gamma}={\varepsilon_{\alpha}}^{\mu\gamma}\rho_{\mu}$,
y $\rho_{\mu}=\frac{\partial_{\mu}}{\Box^{\frac{1}{2}}}$.
Realizando al operador $\mathbb{P}_{\mu}$ como
$i\partial_{\mu}$,(\ref{eq:cap23}) se escribe como

\begin{equation}
h_{\alpha\beta}^{\pm Tt} =
\frac{1}{2}[{\mathbb{I}_{{\alpha\beta}}}^{\gamma\sigma} \mp
\frac{1}{2\Box^{1/2}}({\mathbb{P}_{\mu}\mathbb{J}^{\mu})_{\alpha\beta}}
^{\gamma\sigma}]h_{\alpha\sigma}^{Tt}.
\end{equation}
As\'i

\begin{equation}\label{eq:cap25}
h_{\alpha\beta}^{Tt}=h_{\alpha\beta}^{+Tt} + h_{\alpha\beta}^{-Tt}
\end{equation}

\begin{equation}\label{eq:cap26}
[\mathbb{P}_{\mu}\mathbb{J}^{\mu}h^{Tt}]_{\alpha\beta}
=2\Box^{1/2}(h_{\alpha\beta}^{-Tt} - h_{\alpha\beta}^{+Tt})
\end{equation}
Vamos, entonces a (\ref{eq:cap22}) con (\ref{eq:cap25}) y
(\ref{eq:cap26}), obteniendo

\begin{equation}
-(\Box^{1/2} \mp m)h_{\mu\nu}^{+Tt} - (\Box^{1/2} \pm
m)h_{\mu\nu}^{-Tt}=0.
\end{equation}
Vemos de esta forma que en la capa de masa $(\Box^{1/2} \sim \vert
m\vert)$ una de las dos componentes irreducibles tiene
propagaci\'on dependiendo del signo de $m$.

Volviendo a la ecuaci\'on (\ref{eq:cap22}), en componentes queda
como

\begin{equation}\label{eq:cap28}
\mp
{\varepsilon_{\alpha}}^{\mu\gamma}\partial_{\mu}h_{\gamma\beta}^{Tt}
+ mh_{\alpha\beta}^{Tt}=0.
\end{equation}

Teniendo en cuenta la expresi\'on de la conexi\'on de spin de
torsi\'on nula cuando $h_{\mu\nu}=h_{\mu\nu}^{Tt}$ (ver el
Ap\'endice)

\begin{equation}
\omega_{\mu\nu}(h^{Tt})={\varepsilon_{\mu}}
^{\sigma\lambda}\partial_{\sigma}h_{\lambda\nu}^{Tt},
\end{equation}
as\'i (\ref{eq:cap28}) puede escribirse como

\begin{equation}
h_{\alpha\beta}^{Tt}= \pm \frac{1}{m}\omega_{\alpha\beta}(h^{Tt}),
\end{equation}
que es como una ``condici\'on de autodualidad'' en
$h_{\alpha\beta}^{Tt}$ y la conexi\'on $\omega_{\alpha\beta}$ que
generaliza la que aparece en el caso vectorial.

La acci\'on que nos lleva a la ecuaci\'on (\ref{eq:cap28}) luego
de eliminar los grados de libertad esp\'urios es la acci\'on
autodual (SD)~\cite{ArKh}\cite{deserself}

\begin{eqnarray}\label{eq:cap211}
S_{SD}^{\pm} & = &-\frac{ms}{4}\int \rd^{3}x\,[h_{\mu\rho}
\varepsilon^{\mu\nu\lambda}\partial_{\nu}{h_{\lambda}}^{\rho}+
\frac{ms}{2}(h_{\mu\nu}h^{\nu\mu} -h_{\mu}^{\mu}h_{\nu}^{\nu})],\ \ \ \ \textrm{(a)} \nonumber \\
& = & -\frac{ms}{4}\int \rd^{3}x\, {h_{\mu}}^{a}[
\varepsilon^{\mu\nu\lambda}\eta_{ab}
\partial_{\nu}+\frac{ms}{2}\varepsilon_{abc}\varepsilon^{\mu\lambda\rho}\delta_{\rho}^{c}]
{h_{\lambda}}^{b},\ \ \ \ \ \ \ \ \  \ \textrm{(b)} \\
& \equiv & \frac{m}{2}\int \rd^{3}x\, {h_{\mu}}^{a}{{{{K^{\pm}}^
{\mu}}_{a}}^{\lambda}}_{b} {h_{\lambda}}^{b},\qquad \qquad \ \ \ \
\qquad \ \ \ \ \ \ \ \ \ \ \ \ \ \ \ \ \textrm{(c)} \nonumber
\end{eqnarray}
con $s=\pm 2$. En (\ref{eq:cap211}b) se expresa la acci\'on
autodual de esa forma para recordar que estamos pensando a
${h_{\mu}}^{a}$ como la linealizaci\'on de la tr\'iada, igualmente
queda claro cual es el operador de evoluci\'on de la acci\'on
autodual

\begin{equation}
{{{{K^{\pm}}^ {\mu}}_{a}}^{\lambda}}_{b}=-\frac{s}{2}(
\varepsilon^{\mu\nu\lambda}\eta_{ab}\partial_{\nu} +
\frac{ms}{2}\varepsilon_{abc}\varepsilon^{\mu\lambda\rho}\delta_{\rho}^{c}).
\end{equation}

El campo $h_{\mu\nu}$ en (\ref{eq:cap211}) no tiene simetr\'ias.
Uno podr\'ia pensar que, dado que la ecuaci\'on (\ref{eq:cap28})
s\'olo involucra la parte sim\'etrica, que la parte
antisim\'etrica de  $h_{\mu\nu}$ tiene poco que ver, lo cual no es
cierto. Para ver esto hagamos la descomposici\'on

$$h_{\mu\nu}=H_{\mu\nu} + \varepsilon_{\mu\nu\lambda}V^{\lambda}
\ \textrm{con} \ H_{\mu\nu}=H_{\nu\mu} $$


\begin{eqnarray}
S_{SD}^{\pm} & = & \frac{m}{2}\int \rd^{3}x\,[\mp H_{\mu\rho}\varepsilon^{\mu\nu\lambda}\partial_{\nu}H_{\lambda}^{\rho}- m(H_{\mu\nu}H^{\mu\nu} -H_{\mu}^{\nu}H_{\nu}^{\nu}) +\nonumber  \\
& & \pm 2 V^{\mu}(\partial_{\rho}H_{\mu}^{\rho} -
\partial_{\mu}H_{\rho}^{\rho}) \pm
\varepsilon^{\mu\nu\lambda}V_{\mu}\partial_{\nu}V_{\lambda} -
2mV_{\mu}V^{\mu}],
\end{eqnarray}
donde observamos que la parte antisim\'etrica aparece en su parte
cuadr\'atica como una exitaci\'on autodual de masa $2\vert m\vert$
y que adem\'as interact\'ua con la parte de spin 1 del $H_{\mu\nu}
(\sim \partial^{\mu}H_{\mu\nu})$ la cual se propagar\'ia con masa
$2\vert m\vert$ si no estuviera $V_{\mu}$. El campo $V_{\mu}$
aparece entonces como un campo auxiliar cuya funci\'on es cancelar
la propagaci\'on de la parte de spin 1 de $H_{\mu\nu}$. Las partes
de spin 0 tampoco tienen propagaci\'on alguna dej\'andonos solo
con la parte $h_{\mu\nu}^{Tt}$ de $h_{\mu\nu}$.

Se puede realizar el an\'alisis can\'onico de la teor\'ia SD
corrobor\'andose que hay un s\'olo grado de libertad y de que el
hamiltoniano asociado es definido
positivo~\cite{ArKh}\cite{PJA}\cite{piorolando}, independiente del
signo $\pm$ que solo tiene que ver con el spin de la excitaci\'on.
\'Este an\'alisis se hace por la v\'ia de la acci\'on reducida y
adem\'as permite verificar que la teor\'ia es covariante on shell
y que el spin es $s=2\frac{m}{\vert m \vert}=\pm
2$~\cite{piorolando}.

Tanto la condici\'on de Pauli-Lubanski (\ref{eq:cap22}) como su
expresi\'on (\ref{eq:cap28}), nos sugieren que podr\'iamos pensar
en su ecuaci\'on de movimiento sobre los grados f\'isicos en
funci\'on de un tensor que sea 2-covariante, sim\'etrico,
transverso y sin traza. En este punto traemos a colaci\'on el
hecho que para la conexi\'on de spin lienalizada y el tensor de
Einstein linealizado sucede que (ver el Ap\'endice)

\begin{eqnarray}
\omega_{\mu\nu}(h^{Tt}) & = & \varepsilon_{\mu}^{\sigma\lambda}\partial_{\sigma}H_{\lambda\nu}^{Tt}\nonumber\\
& = & W_{\mu\nu}^{Tt},
\end{eqnarray}

\begin{eqnarray}
G_{\mu\nu}(h^{Tt}) & = & - \Box H_{\mu\nu}^{Tt} \nonumber\\
& = &G_{\mu\nu}^{Tt}.
\end{eqnarray}
Tendremos, entonces, dos ecuaciones mas asociadas a teor\'ias de
spin 2.

\begin{eqnarray}\label{eq:cap216}
[{\mathbb{J}^{\mu} \mathbb{P}_{\mu} \pm 2m
\mathbb{I}]_{\alpha\beta}}^
{\gamma\sigma}W_{\gamma\sigma}^{Tt}(h^{Tt}) & = & 0,
 \ \ \ \ \textrm{(a)} \nonumber \\
{\lbrack \mathbb{J}^{\mu} \mathbb{P}_{\mu} \pm 2m
\mathbb{I}\rbrack_{\alpha\beta}}^{\gamma\sigma}G_{\gamma\sigma}^{Tt}(h^{Tt})
& = & 0 \ \ \ \ \textrm{(b)}.
\end{eqnarray}
En el primer caso las ecuaciones de segundo orden corresponden a
la teor\'ia intermedia (TI)~\cite{ArKh} y en el segundo caso el
sistema de ecuaciones de tercer orden corresponden a la teor\'ia
topol\'ogica masiva (TM) linealizada~\cite{DJT}. Ambos modelos se
pueden estudiar de forma can\'onica y se prueba que describen un
s\'olo grado de libertad y que el spin es $s=2\frac{m}{\vert m
\vert}$~\cite{ArKh}~\cite{DJT}~\cite{PJA}~\cite{jorgepio}.

Recordando la forma en que se escribe la acci\'on SD y la forma de
la conexi\'on de torsi\'on nula (ver el Ap\'endice)

\begin{eqnarray}
\omega_{\mu}^{a} (h) & = & W_{\mu}^{a} (h)\nonumber\\
& = &
\frac{1}{2}\delta_{\lambda}^{a}\varepsilon^{\lambda\nu\gamma}[\partial_{\nu}(h_{\mu\gamma}
+ h_{\gamma\mu}) - \partial_{\mu}h_{\nu\gamma}]
\end{eqnarray}
estas acciones (de la TI y la TM) se pueden escribir como

\begin{eqnarray}\label{eq:cap218}
S^{\pm}_{int} & = & \pm\frac{m}{2}\int \rd^{3}x\, {h_{\mu}}^{a}
{{{{K^{\pm}}^ {\mu}}_{a}}^{\lambda}}_{b}
\frac{1}{m}{W_{\lambda}}^{b}(h)\ \ \ \ \textrm{(a)} \nonumber \\
& = & S_{E}\pm \frac{m}{2}\int \rd^{3}x\,
{h_{\mu}}^{a}\varepsilon^{\mu\nu\lambda}\partial_{\nu}h_{\lambda
a}\ \ \ \ \ \ \   \ \textrm{(b)}
\end{eqnarray}

\begin{eqnarray}\label{eq:cap219}
S^{\pm}_{TM} & = & \frac{m}{2}\int \rd^{3}x\,\,
\frac{1}{m}{W_{\mu}}^{a}(h) {{{{K^{\pm}}^
{\mu}}_{a}}^{\lambda}}_{b}\frac{1}{m}{W_{\lambda}}^{b}(h),
\ \ \ \ \textrm{(a)} \nonumber \\
& = & - S_{E}\mp \frac{m}{2}\int \rd^{3}x\,\,
\frac{1}{2m}{W_{\mu}}^{a}(h)\varepsilon^{\mu\nu\lambda}\partial_{\nu}{W_{\lambda}}^{b}
(\eta)\eta_{ab}.\ \ \ \ \textrm{(b)}
\end{eqnarray}
El \'ultimo t\'ermino en (\ref{eq:cap219}b) se desarrolla como

\begin{eqnarray}
\frac{1}{2}\int \rd^{3}x\,\,
{W_{\mu}}^{a}(h)\varepsilon^{\mu\nu\lambda}\partial_{\nu}W_{\lambda
a}(h)
& = & \frac{1}{2}\int \rd^{3}x\, W_{\mu}^{a}(h)G^{\mu\nu}(h)\eta_{\nu a}, \nonumber \\
& = & -\frac{1}{2}\int \rd^{3}x\,\, {h_{\mu}}^{a}C^{\mu\nu}(h)\eta_{\nu a},\nonumber \\
& \equiv &  S_{c},
\end{eqnarray}
donde $C^{\mu\nu}(h)$ es el tensor de Cotton linealizado.

La acci\'on de Einstein en (\ref{eq:cap218}) y (\ref{eq:cap219})
se escribe en cualquiera de las formas

\begin{eqnarray}
S_{E} & = & -\frac{1}{2}\int
\rd^{3}x\,{h_{\mu}}^{a}G^{\mu\nu}(h)\eta_{\nu a},
\ \ \ \ \ \ \ \qquad \  \textrm{(a)} \nonumber \\
& = &  \frac{1}{2}\int
\rd^{3}x\,{h_{\mu}}^{a}\eta_{ab}\varepsilon^{\mu\nu\lambda}
\partial_{\nu}{W_{\lambda}}^{b}(h),\ \ \ \ \ \ \ \ \ \textrm{(b)}\\
& = & \frac{1}{2}\int
\rd^{3}x\,\varepsilon_{abc}\varepsilon^{\mu\nu\lambda}
\delta_{\lambda}^{c}{W_{\mu}}^{b}(h){W_{\lambda}}^{c}(h).\ \ \ \
\textrm{(c)}\nonumber
\end{eqnarray}

Para resumir lo obtenido hasta ahora con las acciones de spin 2
involucradas y que realizan la condici\'on de Pauli-Lubanski
definimos:

\begin{eqnarray}
S_{TCS} & = & \frac{1}{2}\int
\rd^{3}x\,{h_{\mu}}^{a}\varepsilon^{\mu\nu\lambda}
\partial_{\nu}{h_{\lambda}}^{b}\eta_{ab}\ \ \ \ \ \ \textrm{(a)}\nonumber \\
S_{FP} & = & - \frac{1}{2}\int
\rd^{3}x\,\varepsilon_{abc}\varepsilon^{\mu\nu\lambda}{h_{\mu}}^{a}{h_{\nu}}^{b}\delta_{\lambda}^{c}\
\ \ \textrm{(b)}
\end{eqnarray}
cuyos nombres vienen del t\'ermino tri\'adico de Chern-Simons (o
tambi\'en llamado t\'ermino traslacional de
Chern-Simons)~\cite{VCS} y del t\'ermino tipo Fierz-Pauli,
respectivamente. Con estas definiciones tenemos que las acciones
de spin 2 masivo a considerar se escriben como

\begin{eqnarray}\label{eq:cap222}
S_{SD}^{\pm} & = & \mp mS_{TCS} +m^{2}S_{FP},\ \ \ \ \textrm{(a)} \nonumber \\
S_{int}^{\pm} & = & \pm  mS_{TCS} +S_{E},\ \ \ \ \ \ \ \ \ \textrm{(b)}\\
S_{TM}^{\pm} & = & \mp \frac{1}{m}S_{C} - S_{E},\ \ \ \ \  \
\qquad \textrm{(c)}\nonumber
\end{eqnarray}
donde el signo $\pm$ no influye en la positividad de la energ\'ia,
pero si en el spin. El signo fijo que aparece en cada acci\'on si
tiene que ser asi para que la energ\'ia sea definida positiva. Es
notorio que en la acci\'on $S_{TM}$ el signo del t\'ermino de
Einstein es opuesto al usual.

La acci\'on $S_{int}^{\pm}$ que realiza la condici\'on
(\ref{eq:cap216}a) es invariante, salvo un t\'ermino de borde,
bajo transformaciones de difeomorfismos linealizadas, que ser\'a
una invariancia de calibre de la teor\'ia. An\'alogamente la
ecuaciones de movimiento de (\ref{eq:cap216}c) adem\'as de ser
invariante bajo difeomorfismos, tambi\'en lo son bajo
transformaciones de Lorentz, as\'i que $S_{TM}$ tiene estas dos
invariancias de calibre. De hecho la acci\'on $S_{TM}$ s\'olo
depende de la parte sim\'etrica de $h_{\mu\nu}$. Veremos que en el
proceso de dualidad lo que haremos es ir agregando invariancias de
calibre y aumentando el espacio de soluciones incorporando
soluciones no triviales de $\omega_{\mu\nu}(h)=0$ (en el caso de
la TI) y de $G_{\mu\nu}(h)=0$ (en el caso de la teor\'ia TM). En
sentido contrario uno podr\'ia ir fijando calibre y pasar de la
teor\'ia TM a la TI y de ah\'i a la teor\'ia SD. Esto fu\'e
realizado hace un tiempo en~\cite{jorgepio}.

Una propiedad que comparten todas estas acciones es que la
presencia de los t\'erminos con la densidad de Levy-Civita son
sensibles a las transformaciones discretas P y T. De hecho al
hacer alguna de estas transformaciones pasamos a una teor\'ia del
mismo tipo pero que describe el spin opuesto. Cabe aca preguntar
si es posible, en analog\'ia con el caso vectorial, construir una
teor\'ia que describa excitaciones de spin $2$ y que sea
invariante bajo P y T. La respuesta es que si y para verlo tomemos
el modelo de Einstein masivo descrito por
\begin{equation}
S  = S_{E} +m^{2}S_{FP},
\end{equation}
el cual a primer orden se escribe como
\begin{equation}
S = \frac{1}{2}\int \rd^{3}x\,[2{h_{\mu}^a}
\varepsilon^{\mu\nu\lambda}\partial_{\nu}{w_{\lambda}}^{b}\eta_{ab}
- \varepsilon_{abc}\varepsilon^{\mu\lambda\rho}\delta_{\rho}^{c}
(m^{2}{h_{\mu}}^{a}{h_{\lambda}}^{b} +
{w_{\mu}}^{a}{w_{\lambda}}^{b})], \label{S}
\end{equation}
donde queda claro que el campo $w$ aparece de forma auxiliar como
un pseudo tensor de segundo orden para garantizar la invariancia
bajo las transformaciones impropias P y T. Al hacer variaciones
respecto a \'el obtendremos que es la conexi\'on de spin de
torsi\'on nula linealizada, que al sustituirla nos lleva a la
acci\'on original de segundo orden. Si hacemos, en analog\'ia con
el caso vectorial, el cambio
\begin{eqnarray}
{h_{\mu}}^{a} & = & \frac{1}{\sqrt{2}} ({h_{1\mu}}^{a} + {h_{2\mu}}^{b}), \ \ \textrm{(a)} \nonumber \\
{w_{\mu}}^{a} & = & \frac{m}{\sqrt{2}} ({h_{1\mu}}^{a} -
{h_{2\mu}}^{b}),  \ \ \textrm{(b)}
\end{eqnarray}
 y sustituimos en (\ref{S}) la acci\'on se desacopla en dos modelos autoduales con igual
 masa ($m$) y spines opuestos
\begin{equation}
S \longrightarrow S_{SD}^{-}[h_1] + S_{SD}^{+}[h_2].
\end{equation}
Ahora, si al hacer una de estas transformaciones discretas
inclu\'imos el cambio entre los campos ${h_{1\mu}}^{a}$ y
${h_{2\mu}}^{a}$ \'esta permanece invariante.

\section{Dualidad $S_{SD}\longrightarrow S_{int}$}

En esta secci\'on generalizaremos el procedimiento llevado a cabo
con los modelos vectoriales~\cite{jorge}\cite{pioroma}. Partimos
de la acci\'on (\ref{eq:cap222})

\begin{equation}
S_{SD}=-\frac{ms}{4}\int
\rd^{3}x\,(h_{\mu}^{a}\varepsilon^{\mu\nu\lambda}\partial_{\nu}h_{\lambda
a} +
\frac{ms}{2}\varepsilon_{abc}\varepsilon^{\mu\nu\lambda}h_{\mu}^{a}h_{\nu}^{b}\delta_{\lambda}^{c}).\label{SD0}
\end{equation}

La funci\'on de partici\'on de esta acci\'on es

\begin{equation}
Z_{SD}(0)=N\int\, \mathcal{D}h_{\mu\nu}e^{-S_{SD}}.
\end{equation}

El primer t\'ermino en (\ref{SD0}) es invariante, salvo un
t\'ermino de borde, bajo la suma a $h_{\mu}^{a}$ de
$\partial_{\mu}b^{a}$. As\'i le sumamos el campo auxiliar $H_{\mu
a}$

\begin{equation}
-\frac{ms}{4}\int \rd^{3}x\,(h_{\mu}^{a} +
H_{\mu}^{a})\varepsilon^{\mu\nu\lambda}\partial_{\nu}(h_{\lambda
a} + H_{\lambda a}),
\end{equation}
e imponemos la condici\'on

\begin{equation}\label{eq:cap225}
\varepsilon^{\mu\nu\lambda}\partial_{\nu}H_{\lambda a}=0,
\end{equation}
que nos permite realizar la transformaci\'on de dualidad. De hecho
(\ref{eq:cap225}) es equivalente a tomar ${W_{\mu}}^{a}(H)=0$.
Viendo la forma de ${W_{\mu}}^{a}(H)$ obtendremos que
(\ref{eq:cap225}) obliga a que localmente $H_{\lambda
a}=\partial_{\lambda}b_{a}$, que es calibrable v\'ia
transformaciones de difeomorfismos. Tenemos que la soluci\'on de
(\ref{eq:cap225}) es puro calibre y se recupera $S_{SD}$
(\ref{eq:cap222}). En general para incluir soluciones generales de
$W_{\mu}^{a}(H)=0$, se forza (\ref{eq:cap225}) con un
multiplicador de Lagrange. La acci\'on a considerar es entonces

\begin{eqnarray}\label{eq:cap226}
\widetilde{S}_{SD} & = & -\frac{ms}{4}\int \rd^{3}x\,
[(h_{\mu}^{a} + H_{\mu}^{a})\varepsilon^{\mu\nu\lambda}\partial_{\nu}(h_{\lambda a} + H_{\lambda a}) + \nonumber\\
& & +
\frac{ms}{2}\varepsilon_{abc}\varepsilon^{\mu\nu\lambda}h_{\mu}^{a}h_{\nu}^{b}\delta_{\lambda}^{c}
- 2B_{\mu a} \varepsilon^{\mu\nu\lambda} \partial_{\nu} H_{\lambda
a}^{a}],
\end{eqnarray}
si en el \'ultimo t\'ermino hacemos el cambio

\begin{equation}
B_{\mu a}\longrightarrow \lambda_{\mu a}-\frac{1}{2}\eta_{\mu
a}\lambda_{\mu}^{\mu},
\end{equation}
\'este toma la forma $\lambda_{\mu a}W^{\mu a}(H)$. En la integral
funcional \'este t\'ermino promueve una $\delta({W_{\mu}}^{a}(H))$
posiblemente debiendo tener en cuenta algunas sutilezas de
caracter topol\'ogico~\cite{marioalvaro} que no pretendemos
abordar en este trabajo. Podremos as\'i recuperar $S_{AD}$ a nivel
de la acci\'on o de su integral funcional. El proceso de dualidad
que desarrollaremos lo que pretende es ver cual es la acci\'on que
resulta luego de eliminar los campos y dejar la acci\'on que
resulte para el multiplicador $B_{\mu a}$. La acci\'on que quede
ser\'a la del modelo dual.

Veamos el proceso a nivel de la acci\'on. Las ecuaciones de
movimiento (\ref{eq:cap226}) son:

\begin{eqnarray}\label{eq:cap228}
\frac{\delta\widetilde{S}}{\delta h_{\mu a }} & = & 0
\Longrightarrow
-\varepsilon^{\mu\nu\lambda}\partial_{\nu}({h_{\lambda}}^{a}+
{H_{\lambda}}^{a})
-\frac{ms}{2}{\varepsilon^{a}}_{bc}\varepsilon^{\mu\nu\lambda}{h_{\nu}}^{b}\delta_{\lambda}^{c}=0, \ \textrm{(a)} \nonumber \\
\frac{\delta\widetilde{S}}{\delta H_{\mu a }} & = & 0
\Longrightarrow
-\varepsilon^{\mu\nu\lambda}\partial_{\nu}({h_{\lambda}}^{a}+
{H_{\lambda}}^{a}
 - {B_{\lambda}}^{a})=0 \qquad \qquad \ \ \textrm{(b)},\nonumber \\
\frac{\delta\widetilde{S}}{\delta B_{\mu a}} & = & 0 \qquad
\varepsilon^{\mu\nu\lambda}\partial_{\nu}{H_{\lambda}}^{a}=0. \ \
\ \ \qquad \qquad\qquad \qquad \qquad \textrm{(c)}
\end{eqnarray}
Las ecuaciones de movimiento son invariantes bajo las
transformaciones de calibre

\begin{eqnarray}
\delta H_{\lambda}^{a}= \partial_{\lambda}\zeta^{a}\textrm{(a)}, \nonumber \\
\delta B_{\lambda}^{a}=
\partial_{\lambda}\tilde{\zeta}^{a}.\textrm{(b)}
\end{eqnarray}

La ecuaci\'on  (\ref{eq:cap228}c) puede resolverse localmente
permitiendo calibrar $H_{\lambda}^{a}$ y al ir a (\ref{eq:cap226})
recuperamos (\ref{eq:cap222}). Para que emerja la teor\'ia dual
observamos que de (\ref{eq:cap228}b)

\begin{eqnarray}\label{eq:cap2300}
{H_{\lambda}}^{a}={B_{\lambda}}^{a}-{h_{\lambda}}^{a}+{l_{\lambda}}^{a}\ \textrm{(a)}, \nonumber \\
\textrm{con}  \  \ \ \ \
\varepsilon^{\mu\nu\lambda}\partial_{\nu}{l_{\lambda}}^{a}=0, \
\textrm{(b)}
\end{eqnarray}
que sustituimos en (\ref{eq:cap226})

\begin{eqnarray}\label{eq:cap231}
\widetilde{\widetilde{S}}_{SD}=\widetilde{S}_{SD}\Big\vert_{(37)}
& = & -\frac{ms}{4}\int \rd^{3}x\,
[{B_{\mu}}^{a}\varepsilon^{\mu\nu\lambda}\partial_{\nu}B_{\lambda
a} + \frac{ms}{2}
\varepsilon_{abc}\varepsilon^{\mu\nu\lambda}{h_{\mu}}^{a}{h_{\nu}}^{b}\delta_{\lambda}^{c}
+ \nonumber \\
 & & -2{B_{\mu}}^{a}\varepsilon^{\mu\nu\lambda} \partial_{\nu}(B_{\lambda a} - h_{\lambda a})], \nonumber \\
& = & -\frac{ms}{4} \int \rd^{3} x
[-{B_{\mu}}^{a}\varepsilon^{\mu\nu\lambda}
\partial_{\nu} B_{\lambda a} + \frac{ms}{2} \varepsilon_{abc}\varepsilon^{\mu\nu\lambda}
{h_{\mu}}^{a}{h_{\nu}}^{b}\delta_{\lambda}^{c}+ \nonumber \\
& & +2{B_{\mu}}^{a}\varepsilon^{\mu\nu\lambda} \partial_{\nu}
h_{\lambda a}].
\end{eqnarray}
Que corresponde a una versi\'on a primer orden de $S_{int}$. De
hecho, si en $\widetilde{\widetilde{S}}_{SD}$ hacemos variaciones
respecto a $h_{\mu a}$

\begin{equation}
\frac{\delta\widetilde{\widetilde{S}}_{SD}}{\delta h_{\mu a}}=0
\Longrightarrow
\varepsilon^{\mu\nu\lambda}\partial_{\nu}{B_{\lambda}}^{a}=-\frac{ms}{2}{\varepsilon_{bc}}^{a}
\varepsilon^{\mu\nu\lambda}{h_{\nu}}^{b}\delta_{\lambda}^{c}
\end{equation}
y as\'i (ver el Ap\'endice)

\begin{eqnarray}
{h_{\mu}}^{a} & = & -\frac{2}{ms}{{{W_{\mu}}^{a}}^{\lambda}}_{b}{B_{\lambda}}^{b}\nonumber \\
& = & -\frac{2}{ms}{W_{\mu}}^{a}(B)
\end{eqnarray}
que nos permite eliminar a ${h_{\mu}}^{a}$, qued\'andonos el
modelo dual

\begin{eqnarray}
S^{dual}_{SD} & = & -\frac{ms}{4}\int \rd^{3} x
[-{B_{\mu}}^{a}\varepsilon^{\mu\nu\lambda}\partial_{\nu}
B_{\lambda a} -
\frac{2}{ms}\varepsilon_{abc} \varepsilon^{\mu\nu\lambda}{W_{\mu}}^{a}(B){W_{\nu}}^{b}(B)\delta_{\lambda}^{c}+\nonumber \\
& &
+\frac{4}{ms}{B_{\mu}}^{a}\varepsilon^{\mu\nu\lambda}\partial_{\nu}{W_{\lambda}}^{a}(B)],
\nonumber \\
& = & \frac{ms}{4}\int
\rd^{3}x\,{B_{\mu}}^{a}\varepsilon^{\mu\nu\lambda}\partial_{\nu}
B_{\lambda a} + S_{E}(B), \ \ \textrm{(a)} \nonumber \\
& = & S_{int}(B)\qquad \qquad \ \ \ \ \ \ \ \ \ \qquad \ \ \ \ \ \
\  \textrm{(b)}
\end{eqnarray}

Es importante notar que en el modelo dual el t\'ermino TCS aparece
con un signo opuesto al que tiene originalmente en la teor\'ia SD
de partida. Este cambio no est\'a asociado a un cambio de spin,
\'ese es el signo que debe tener para que describa el mismo spin
que el modelo de partida. Notemos que en el c\'alculo se mantuvo
el valor de $s$ sin especificar para que sirva a ambos casos
($s=\pm 2$).

Para conectar las funciones de partici\'on notamos que en
(\ref{eq:cap231}) podemos ``completar cuadrados''

\begin{eqnarray}
-{B_{\mu}}^{a}\varepsilon^{\mu\nu\lambda}\partial_{\nu}B_{\lambda
a}+2{B_{\mu}}^{a}
\varepsilon^{\mu\nu\lambda}\partial_{\nu}h_{\lambda a} & = & -({B_{\mu}}^{a}-{h_{\mu}}^{a})\varepsilon^{\mu\nu\lambda}\partial_{\nu}(B_{\lambda a} - h_{\lambda a}) +\nonumber \\
 & & +{h_{\mu}}^{a}\varepsilon^{\mu\nu\lambda}\partial_{\nu}h_{\lambda a}
\end{eqnarray}
as\'i

\begin{equation}\label{eq:cap235}
\widetilde{\widetilde{S}}_{SD}=S_{SD}^{+}(h)+S_{TCS}(B-h)
\end{equation}
y entonces

\begin{eqnarray}\label{eq:cap236}
Z_{\widetilde{\widetilde{S}}_{SD}} &\propto& Z_{S_{SD}}Z_{S_{TCS}}\\
&\propto&Z_{S_{int}}
\end{eqnarray}

Haciendo un procedimiento an\'alogo en $\widetilde{S}_{SD}$
obtendremos
\begin{equation}\label{eq:cap237}
\widetilde{S}_{SD}=\widetilde{\widetilde{S_{SD}}}(h)-\frac{ms}{2}S_{TCS}(H+h-B),
\end{equation}
de donde entendemos que al sustituir (\ref{eq:cap2300}a) el
proceso correspondiente en la integral funcional equivale a hacer
la integral de $\widetilde{H}_{\mu a}=H_{\mu a}+h_{\mu a}-B_{\mu
a}$, incluyendo los t\'erminos de fijaci\'on de calibre,
produciendo un factor $Z_{S_{TCS}}$.

Observamos en (\ref{eq:cap236}) que las funciones de partici\'on
del modelo dual $(S_{int})$ y el modelo SD diferiran en un factor
asociado a la funci\'on de partici\'on del modelo TCS, el cual
est\'a asociado a soluciones de ${\omega_{\mu}}^{a}(h)=0$. El
significado topol\'ogico o no trivial que pueda tener este factor
merece ser estudiado. \'Este podr\'ia cobrar valor si consideramos
variedades con topolog\'ia no trivial\footnote{La dualidad vista
desde el punto de vista de las funciones de partición ha sido
considerada recientemente de forma mas completa e independiente
en\cite{dalmazi}}.

El procedimiento que seguimos incorpor\'o la invariancia de
calibre bajo difeomorfismos y ya es conocido que $S_{int}$ y
$S_{SD}$ se conectan fijando el calibre en el
primero~\cite{jorgepio}. Esta situaci\'on es an\'aloga al caso
vectorial. Pasamos ahora a la dualidad entre el TI y la TM donde
incorporaremos la invariancia que falta: la invariancia bajo
transformaciones de Lorentz linealizadas.

\section{Dualidad $S_{int}\longrightarrow S_{TM}$}

Partimos, ahora, de la acci\'on intermedia con el t\'ermino de
Einstein escrito a segundo orden orden

\begin{equation}
S_{int}=\frac{1}{2}\int
\rd^{3}x\,[{h_{\mu}}^{a}\varepsilon^{\mu\nu\lambda}
\partial_{\nu}\omega_{\lambda a}(h)+
\frac{ms}{2}{h_{\mu}}^{a}\varepsilon^{\mu\nu\lambda}\partial_{\nu}h_{\lambda
a}]. \label{sint}
\end{equation}
Esta acci\'on es invariante, salvo un t\'ermino de borde, bajo

\begin{equation}
\delta h_{\mu}^{a}=\partial_{\mu}\zeta^{a}
\end{equation}
que corresponde a la invariancia bajo difeomorfismos linealizada.
Podemos observar tambi\'en que la parte que corresponde al
t\'ermino de Einstein es invariante, salvo un t\'ermino de borde,
bajo la transformaciones de Lorentz linealizadas
\begin{eqnarray}
\delta
{h_{\mu}}^a={\varepsilon_{\mu}}^{a\sigma}l_{\sigma},&&\delta
{\omega_{\mu}}^{a}(h)=-\partial_{\mu}l^{a}.
\end{eqnarray}
\'Esta invariancia no la tiene el t\'ermino TCS. Para promoverla
modificamos el t\'ermino en forma an\'aloga a antes
\begin{equation}
-\frac{ms}{4}\int \rd^{3}x\,(h_{\mu}^{a} +
H_{\mu}^{a})\varepsilon^{\mu\nu\lambda}\partial_{\nu}(h_{\lambda
a} + H_{\lambda a}), \label{lorentz1}
\end{equation}
e imponemos, ahora, la condici\'on

\begin{equation}
\varepsilon^{\mu\nu\lambda}\partial_{\nu}W_{\lambda
a}(H)=-{G^{\mu}}_a=0, \label{g0}
\end{equation}
con $W_{\lambda a}(H)=(\eta_{\mu a}\eta_{\lambda
b}-\frac{1}{2}\eta_{\lambda a}\eta_{\mu
b})\varepsilon^{\mu\rho\sigma}\partial_{\rho}{H_{\sigma}}^b$, que
corresponde a la conexi\'on de spin linealizada de torsi\'on nula.
Localmente la soluci\'on de (\ref{g0}) es ``puro calibre'' y
permite recuperar el t\'ermino $TCS$ y as\'i la acci\'on
intermedia. Globalmente \'este no ser\'a el caso.

Empezamos as\'i con la acci\'on
\begin{eqnarray}
\widetilde{S}_{int} & = & \frac{1}{2}\int
\rd^{3}x\,[{h_{\mu}}^{a}\varepsilon^{\mu\nu\lambda}
\partial_{\nu}\omega_{\lambda a}(h)+ \frac{ms}{2}({h_{\mu}}^{a} +
{H_{\mu}}^{a})\varepsilon^{\mu\nu\lambda}\partial_{\nu}(h_{\lambda
a} + H_{\lambda a}) \nonumber \\
&& \qquad \qquad+
2{B_{\mu}}^a\varepsilon^{\mu\nu\lambda}\partial_{\nu}W_{\lambda
a}(H)], \label{sinttilde}
\end{eqnarray}
donde hemos introducido a un multiplicador ${B_{\mu}}^a$ que forza
la condici\'on (\ref{g0}).

Las ecuaciones de movimiento de (\ref{sinttilde}) son
\begin{eqnarray}
\frac{\delta\widetilde{S}}{\delta h_{\mu a }} & = & 0
\Longrightarrow
\varepsilon^{\mu\nu\lambda}\partial_{\nu}[{{\omega}_{\lambda}}^a(h)+\frac{ms}{2}({h_{\lambda}}^{a}+
{H_{\lambda}}^{a})]=0, \ \qquad \textrm{(a)} \nonumber \\
\frac{\delta\widetilde{S}}{\delta H_{\mu a }} & = & 0
\Longrightarrow
\varepsilon^{\mu\nu\lambda}\partial_{\nu}[{W_{\lambda}}^{a}(B)+\frac{ms}{2}({h_{\lambda}}^{a}+
{H_{\lambda}}^{a})]=0. \ \ \textrm{(b)}\nonumber \\
 \frac{\delta\widetilde{S}}{\delta B_{\mu a}} & = & 0 \Longrightarrow \qquad
\varepsilon^{\mu\nu\lambda}\partial_{\nu}{W_{\lambda}}^{a}(H)=0, \
\ \ \ \qquad \qquad \quad \textrm{(c)} \label{ecmov}
\end{eqnarray}
\'Estas son invariantes bajo las transformaciones de calibre

\begin{eqnarray}
\delta h_{\lambda}^{a}= \partial_{\lambda}\xi_1^{a},&\qquad \delta
H_{\lambda}^{a}= \partial_{\lambda}\xi_2^{a},&
\qquad B_{\lambda}^{a}= \partial_{\lambda}\xi_3^{a}, \nonumber \\
\delta h_{\lambda}^{a}={\varepsilon_{\lambda}}^{ab}l_b,& \qquad
\delta H_{\lambda}^{a}=-{\varepsilon_{\lambda}}^{ab}l_b,& \qquad
B_{\lambda}^{a}={\varepsilon_{\lambda}}^{ab}\widetilde{l}_b
\end{eqnarray}
donde observamos que ha sido promovida la invariancia bajo
transformaciones de Lorentz ausente en (\ref{sint}). La
invariancia de calibre asociada a $H_{\lambda}^{a}$ y la
resoluci\'n local de (\ref{ecmov}c) permiten recuperar la acci\'on
intermedia. Para obtener el modelo dual primero notamos que
podemos tomar
\begin{eqnarray}\label{eq:cap230}
{H_{\lambda}}^{a}=-{h_{\lambda}}^{a}-\frac{2}{ms}{W_{\lambda}}^{a}(B)+{l_{\lambda}}^{a}\ \textrm{(a)}, \nonumber \\
\textrm{con}  \  \ \ \ \
\varepsilon^{\mu\nu\lambda}\partial_{\nu}{l_{\lambda}}^{a}=0, \
\textrm{(b)}
\end{eqnarray}
que sustituimos en (\ref{sinttilde}), llev\'andonos a
\begin{equation}
\widetilde{\widetilde{S}}_{int} = \frac{1}{2}\int
\rd^{3}x\,[{h_{\mu}}^{a}\varepsilon^{\mu\nu\lambda}
\partial_{\nu}(\omega_{\lambda a}(h)-2W_{\lambda a}(B))
-\frac{2}{ms}{W_{\mu}}^a(B)\varepsilon^{\mu\nu\lambda}\partial_{\nu}W_{\lambda
a}(B)].\label{sint2tilde}
\end{equation}
Ahora hacemos variaciones independientes en (\ref{sint2tilde})
respecto a $h_{\mu a}$ y obtenemos

\begin{equation}
\varepsilon^{\mu\nu\lambda}\partial_{\nu}{W_{\lambda}}^{a}(h-B)=0.
\end{equation}
Usamos la invariancia de calibre Lorentz y de difeomorfismos para
fijar ${h_{\mu}}^a={B_{\mu}}^a$ y finalmente sustituyendo esta
fijaci\'on en (\ref{sint2tilde}) obtendremos
\begin{eqnarray}
S_{int}^{dual} &=&
\widetilde{\widetilde{S}}_{int}\Big\vert_{(3.57)}\nonumber \\
&=&-\frac{1}{2}\int \rd^{3}x\,[\varepsilon^{\mu\nu\lambda}
{B_{\mu}}^{a}\partial_{\nu} W_{\lambda a}(B)+
\frac{2}{ms}{W_{\mu}}^a(B)\varepsilon^{\mu\nu\lambda}\partial_{\nu}W_{\lambda
a}(B)], \nonumber\\
&& \label{dualtm}\\
 &=& S_{TM}(B).\nonumber
\end{eqnarray}
Notamos que el primer t\'ermino en (\ref{dualtm}) corresponde a la
acci\'on de Einstein con el signo opuesto al usual, tal como debe
suceder en la acci\'on topo0l\'ogica masiva.

En el proceso cuando pasamos de $\widetilde{S}_{int}$ a
$\widetilde{\widetilde{S}}_{int}$ agotamos la libertad de calibre
asociada a $H_{\mu a}$ en el modelo original. En el pasaje a
$S_{TM}$ agotamos la asociada a $h_{\mu a}$, dej\'andonos
finalmente con la invariancia bajo transformaciones de Lorentz y
de difeomorfismos linealizados en $B_{\mu a}$. De hecho puede
verse que su parte antisim\'etrica no aparece en (\ref{dualtm}).

Para ver la relaci\'on entre las funciones de partici\'on
observemos que $\widetilde{\widetilde{S}}_{int}$ puede escribirse
como
\begin{eqnarray}
\widetilde{\widetilde{S}}_{int} & = & \frac{1}{2}\int
\rd^{3}x\,({h_{\mu}}^{a}-{B_{\mu}}^{a})\varepsilon^{\mu\nu\lambda}
\partial_{\nu}\omega_{\lambda a}(h-B)+ \nonumber \\
& &-\frac{1}{2}\int \rd^{3}x\,[\varepsilon^{\mu\nu\lambda}
{B_{\mu}}^{a}\partial_{\nu} W_{\lambda a}(B)+
\frac{2}{ms}{W_{\mu}}^a(B)\varepsilon^{\mu\nu\lambda}\partial_{\nu}W_{\lambda
a}(B)]. \label{sinttilde2}
\end{eqnarray}
La primera integral en (\ref{sinttilde2}) corresponde a $S_E$ en
el campo ${h_{\mu}}^{a}-{B_{\mu}}^{a}$, la segunda integral es
presisamente $S_{TM}$. As\'i podemos hacer la integral funcional
en $\tilde{h}_{\mu a} \equiv h_{\mu a}-B_{\mu a}$, introduciendo
los t\'erminos de fijaci\'on de calibre que correspondan y
quedar\'ia la integral de la $TM$. As\'i
\begin{equation}
Z_{\widetilde{\widetilde{S}}_{int}} \propto Z_{S_{TM}}Z_{S_{E}}.
\label{Ztildetilde}
\end{equation}
El factor que tiene que ver con la acci\'on de Einstein est\'a en
correspondencia con la fijaci\'on de calibre que se tuvo que hacer
para llegar a la acci\'on $S_{TM}(B)$. En otra direcci\'on
$\widetilde{S}_{int}$ puede reescribirse como
\begin{eqnarray}
\widetilde{S}_{int} & = & \frac{1}{2}\int
\rd^{3}x\,[({h_{\mu}}^{a}+{H_{\mu}}^{a})\varepsilon^{\mu\nu\lambda}
\partial_{\nu}\omega_{\lambda a}(h+H)+ \frac{ms}{2}({h_{\mu}}^{a} +
{H_{\mu}}^{a})\varepsilon^{\mu\nu\lambda}\partial_{\nu}(h_{\lambda
a} + H_{\lambda a}) \nonumber \\
&& \qquad \qquad+
(2{B_{\mu}}^a-2{h_{\mu}}^a-{H_{\mu}}^a\varepsilon^{\mu\nu\lambda}\partial_{\nu}W_{\lambda
a}(H)], \label{tilde3}
\end{eqnarray}
donde vemos como que se han ''aislado´´ los t\'erminos que tienen
que ver con la acci\'on de la $TI$ expresados en funci\'on del
campo $\tilde{h}_{\mu a} \equiv h_{\mu a}+H_{\mu a}$. Al plantear
la integral funcional redefinimos adem\'as $\tilde{B}_{\mu
a}\equiv 2B_{\mu a}-2h_{\mu a}-H_{\mu a}$. La integral funcional
ser\'a la de la acci\'n intermedia por un factor que, al igual que
en el caso de la acci\'on BF del caso vectorial, debe ser
proporcional al cuadrado de la funci\'on de partici\'on de la
acci\'on de Einstein. As\'i
\begin{equation}
Z_{\widetilde{S}_{int}} \propto Z_{S_{int}}(Z_{S_{E}})^2.
\label{Ztilde}
\end{equation}

Teniendo en cuenta (\ref{Ztildetilde}) y (\ref{Ztilde})
conclu\'imos que
\begin{equation}
Z_{\widetilde{S}_{TM}} \propto Z_{S_{int}}Z_{S_{E}},
\label{ZintTM}
\end{equation}
que relaciona las funciones de partici\'on de las acciones
$S_{int}$ y $S_{TM}$.

\section{Conclusi\'on}
Hemos visto que hay tres modelos asociados a la condici\'on de
Pauli-Lubanski con $s=\pm2$. Uno es el modelo autodual ($SD$)
\begin{equation}
S_{SD}^{\pm} = -\frac{ms}{4}\int \rd^{3}x\, {h_{\mu}}^{a}[
\varepsilon^{\mu\nu\lambda}\eta_{ab}
\partial_{\nu}+\frac{ms}{2}\varepsilon_{abc}\varepsilon^{\mu\lambda\rho}\delta_{\rho}^{c}]
{h_{\lambda}}^{b},
\end{equation}
el cual no tiene invariancias de calibre. Le sigue el modelo
intermedio ($TI$)
\begin{equation}
S_{int}=\frac{1}{2}\int
\rd^{3}x\,[{h_{\mu}}^{a}\varepsilon^{\mu\nu\lambda}
\partial_{\nu}\omega_{\lambda a}(h)+
\frac{ms}{2}{h_{\mu}}^{a}\varepsilon^{\mu\nu\lambda}\partial_{\nu}h_{\lambda
a}],
\end{equation}
el cual es invariante, salvo un t\'ermino de borde, bajo las
transformaciones de difeomorfismos linealizadas
\begin{equation}
\delta h_{\mu}^{a}=\partial_{\mu}\zeta^{a}.
\end{equation}
Finalmente est\'a el modelo topol\'ogico masivo ($TM$)
\begin{equation}
S_{TM}= -\frac{1}{2}\int \rd^{3}x\,[\varepsilon^{\mu\nu\lambda}
{h_{\mu}}^{a}\partial_{\nu} \omega_{\lambda a}(h)+
\frac{2}{ms}{\omega_{\mu}}^a(h)\varepsilon^{\mu\nu\lambda}\partial_{\nu}\omega_{\lambda
a}(h)],
\end{equation}
el cual tambi\'en es invariante bajo difiemorfismos y
expl\'icitamente invarianle bajo transformaciones de Lorentz
$\delta h_{\mu a}={\varepsilon_{\mu a}}^b l_b$, dado que la parte
antisim\'etrica de $h_{\mu a}$ no aparece en la acci\'on.

Vimos que se puede pasar por una transformaci\'on de dualidad del
modelo $SD$ al $TI$ incorporando adem\'as la invariancia bajo
difeomorfismos. De forma an\'aloga se pasa del model $TI$ al $TM$
por una transformaci\'on de dualñidad, y se incorpora tambi\'en la
invariancia Lorentz faltante. Los espacios de soluciones de los
modelos $SD$ y $TI$ didieren en las soluciones no triviales de
$\omega_{\mu a}(h)=0$. Estas soluciones corresponden al modelo
$TCS$ libre. En el caso de los modelos $TI$ y $TM$ los espacios de
soluciones difieren en las soluciones no triviales de $G_{\mu
a}(h)=0$, las cuales estan asociadas al modelo no din\'amico de
Einstein libre.

A nivel de las funciones de partici\'on observamos que estas
difieren en factores que estan relacionadas con las funciones de
partici\'on de los modelos $TCS$ y de Einstein. En particular
obtuvimos que
\begin{eqnarray}
Z_{S_{int}}&\propto& Z_{S_{SD}}Z_{S_{TCS}},\\
Z_{S_{TM}}&\propto& Z_{S_{int}}Z_{S_{E}}.
\end{eqnarray}
Notamos que el factor en que difieren las acciones est\'a
relacionado con la parte del espacio de soluciones en que
difieren.

Los c\'aculos aca presentados en relaci\'on con las funciones de
partici\'on se ha hecho de forma heur\'istica y merecen ser
estudiados de forma rigurosa. En varidades de topolog\'ia trivial
los modelos $SD$, $TI$ y $TM$ son completamente equivalentes y sus
funciones de partici\'on seran iguales.

\chapter{Conclusiones}

Siguiendo lineamientos convencionales  se obtienen las ecuaciones
que deben cumplir los campos que realicen el \'algebra del grupo
de Poincar\'e en dimensi\'on 2+1 con spin (o helicidad) entero (no
nulo) y masa m. Se muestra como para el caso de $s=1$ las
realizaciones corresponden al modelo masivo autodual y el
topol\'ogico masivo. Igualmente se muestra que para $s=2$ las
realizaciones corresponden a los modelos: masivo autodual de spin
2, intermedio y topol\'ogico masivo linealizado.

Para $s=1$ se revisa que los modelos estan conectados por ua
transformaci\'on de dualidad la cual incorpora la invariancia de
calibre no presente en el modelo autodual. Se discute tambi\'en la
realci\'on entre las funciones de partici\'on las cuales difieren
en un factor relacionado con la torsi\'on de Ray-Singer de la
variedad base.

Para $s=2$ se muestra que por la v\'ia de transformaciones de
dualidad es posible pasar del modelo autodual al intermedio, y a
partir de \'este \'ultimo al topol\'ogico masivo linealizado. En
el primer paso se incorpora la invariancia bajo difeomorfismos
linealizada ausente en el modelo autodual. En el segundo paso se
incorpora la invariancia bajo transformaciones de Lorentz
linealizada no presente en el modelo intermedio. Para estos
modelos se encuentra que las funciones de partici\'on difieren en
factores asociados a teor\'ias que describen las soluciones no
triviales de ${\omega_{\mu}}^a(h)=0$ (en el caso de los modelos
autodual e intermedio) o a soluciones no triviales de
${G_{\mu}}^a(h)=0$ (en el caso de los modelos intermedio y
topol\'ogico masivo linealizado). Cuando miramos los espacionos de
soluciones de cada par de modelos (autodual-intermedio o
intermedio-topol\'ogico masivo) estos difieren en las soluciones
no triviales antes se\~naladas. Estos factores que diferencian las
funciones de partici\'on cobran importancia a la hora de
considerar las teor\'ias acopladas con fuentes externas. Valdr\'ia
la pena abordar el significado topol\'ogico que podr\'ian tener.

27-12-2009: Recientemente se han publicado resultados,
independientes, concernientes a la dualidad entre las teor\'{\i}as
autodual, intermedia y topol\'ogica masiva desde un enfoque mas
completo de la funci\'on de partici\'on\cite{dalmazi}.

\appendix

\chapter{Convenciones para gravedad curva y linealizada}

\section{Transformaciones bajo difeomorfismos}

Las transformaciones generales de coordenadas

\begin{equation}
x^{\mu}\longrightarrow x^{'\mu}=x^{\mu} - \zeta^{\mu}(x),
\end{equation}

inducen en un tensor $p$ covariante, $q$ contravariante, el cambio

\begin{eqnarray}
\delta T^{\mu_{1}-\mu_{q}}_{\nu_{1}-\nu_{p}} & = & \zeta^{\rho}\partial_{\rho} T^{\mu_{1}-\mu_{q}}_{\nu_{1}-\nu_{p}} + \partial_{\nu_{1}} \zeta^{\rho} T^{\mu_{1}-\mu_{q}}_{\rho \nu_{2}-\nu_{p}}+\\
& & +\partial_{\nu_{2}} \zeta^{\rho} T^{\mu_{1}-\mu_{q}}_{\nu_{1} \rho \nu_{3} - \nu_{p}}+ \cdots \nonumber\\
& & -\partial_{\rho}
\zeta^{\mu_{1}}T^{\rho\mu_{2}-\mu_{q}}_{\nu_{1}-\nu_{p}} -
\partial_{\rho} \zeta^{\mu_{2}}
T^{\mu_{1}\rho-\mu_{q}}_{\nu_{1}-\nu_{p}}-\cdots,  \nonumber
\end{eqnarray}

\section{Derivadas covariantes, conexiones, tensores de Riemman, Ricci, Einstein, Cotton, Torsi\'on}

Las derivadas covariantes mantienen el caracter del objeto sobre el cual act\'uan. Sobre vectores covariantes y contravariantes act\'uan como

\begin{eqnarray}
\mathcal{D}_{\mu}\mathcal{U}_{\nu} & = &\partial_{\mu}\mathcal{U}_{\nu} - \Gamma^{\lambda}_{\mu\nu}\mathcal{U}_{\lambda}, \ \ \ \textrm{(a)}\nonumber \\
\mathcal{D}_{\mu}\mathcal{V}^{\nu} & = &
\partial_{\mu}\mathcal{V}^{\nu} +
\Gamma^{\nu}_{\mu\lambda}\mathcal{V}^{\lambda}.\ \ \  \textrm{(b)}
\end{eqnarray}
Sobre un tensor 2 covariante act\'ua as\'i

\begin{equation}
\mathcal{D}_{\mu}A_{\nu\lambda}=\partial_{\mu}A_{\nu\lambda}-\Gamma^{\rho}_{\mu\nu}A_{\rho\lambda}
- \Gamma^{\rho}_{\mu\lambda}A_{\nu\rho}.
\end{equation}
La generalizaci\'on es inmediata para un tensor mixto.

La conexi\'on $\Gamma^{\lambda}_{\mu\nu}$ no es un buen tensor bajo difeomorfismos, sin embargo la torsi\'on si lo es

\begin{equation}
T^{\lambda}_{\mu\nu}=\Gamma^{\lambda}_{\mu\nu} -
\Gamma^{\lambda}_{\nu\mu}.
\end{equation}

El \emph{tensor de Riemman} se define por el conmutador de las derivadas covariantes

\begin{equation}
[\mathcal{D}_{\mu},\mathcal{D}_{\nu}]\mathcal{V}^{\lambda}={R_{\mu\nu\sigma}}^{\lambda}V^{\sigma}
- T^{\sigma}_{\mu\nu}\mathcal{D}_{\sigma}V^{\lambda},
\end{equation}
donde

\begin{equation}
{R_{\mu\nu\sigma}}^{\lambda}=\partial_{\mu}\Gamma^{\lambda}_{\nu\sigma}
+
\Gamma^{\lambda}_{\mu\rho}\Gamma^{\rho}_{\nu\sigma}-\partial_{\nu}\Gamma^{\lambda}_{\mu\sigma}
+ \Gamma^{\lambda}_{\nu\rho}\Gamma^{\rho}_{\mu\sigma}.
\end{equation}

Cuando la torsi\'on es cero (se dice que el espacio es de Riemman)
se obtiene que

\begin{equation}
\Gamma^{\lambda}_{\mu\nu}=\frac{1}{2}g^{\lambda\sigma}(\partial_{\mu}g_{\sigma\nu}
+ \partial_{\nu}g_{\sigma\mu} - \partial_{\sigma}g_{\mu\nu}).
\end{equation}

Las identidades de Jacobi para los conmutadores nos llevan a las
identidades
\begin{eqnarray}\label{eq:ap29}
\mathcal{D}_{\rho}{R_{\mu\nu\sigma}}^{\lambda} +  T^{\gamma}_{\mu\nu}{R_{\gamma\rho\sigma}}^{\lambda}
+ (\textrm{permutaciones c\'iclicas en} \ \rho\mu\nu) =0. \ \ \ \ \textrm{(a)} \nonumber \\
{R_{\mu\nu\rho}}^{\lambda} - \mathcal{D}_{\rho}
T^{\gamma}_{\mu\nu} - T^{\sigma}_{\mu\nu}T^{\lambda}_{\sigma\rho}
+ (\textrm{permutaciones c\'iclicas en} \ \rho\mu\nu ) =0. \ \ \ \
\textrm{(b)}
\end{eqnarray}

El \emph{tensor de Ricci}  se define por la contracci\'on

\begin{equation}
R_{\mu\nu}={R_{\lambda\mu\nu}}^{\lambda}=R_{\nu\mu},
\end{equation}
y el \emph{escalar de curvatura} se define haciendo otra
contracci\'on

\begin{equation}
R=R^{\mu}_{\mu}.
\end{equation}

Con $R_{\mu\nu}$ y $R$ se define el tensor de Einstein

\begin{equation}
G_{\mu\nu}=R_{\mu\nu} - \frac{1}{2}g_{\mu\nu}R,
\end{equation}
el cual cumple, en virtud de (\ref{eq:ap29})

\begin{equation}
\mathcal{D}_{\mu}G^{\mu}_{\nu}=0
\end{equation}

En $d=2+1$ sucede que los tensores de Einstein y Riemman tienen el mismo n\'umero de componentes independientes. As\'i que uno puede expresarse en funci\'on del otro

\begin{equation}
{R_{\mu\nu\lambda}}^{\sigma}=-\varepsilon_{\mu\nu\rho}{\varepsilon^{\sigma}}_{\gamma\lambda}G^{\rho\lambda}
\end{equation}
y no se tiene tensor de Weyl. En su lugar se usa el tensor de
Cotton

\begin{equation}
C^{\mu\nu}=\frac{1}{\sqrt{-g}}\varepsilon^{\mu\rho\lambda}\mathcal{D}_{\rho}\widetilde{R}^{\nu}_{\lambda}
\end{equation}
con

\begin{eqnarray}
\widetilde{R}^{\nu}_{\mu} & = & G^{\nu}_{\mu} - \frac{1}{2}g^{\nu}_{\mu}G, \ \ \ \  \textrm{(a)} \nonumber \\
& = & R^{\nu}_{\mu} - \frac{1}{4}g^{\nu}_{\mu}R. \ \ \ \
\textrm{(b)}
\end{eqnarray}

\section{El lenguaje de las tr\'iadas y los objetos asociados}

Al remitirnos al espacio tangente se introducen los vielbeins
$e^{a}$ las cuales tiene componentes ${e_{\mu}}^{a}$. Sus duales
tienen componentes ${e^{\mu}}_{a}$. As\'i

\begin{equation}
{e_{\mu}}^{a}{e^{\mu}}_{b}=\delta^{a}_{b}.
\end{equation}

Nos referimos a \'indices de universo con letras griegas y a los
del espacio tangente con las primeras letras del alfabeto.

En el espacio tangente las transformaciones que preservan el producto $U_{a}V^{a}$ son:

\begin{equation}
\delta V^{a} = V^{b} {X_{b}}^{a} \qquad ; \qquad \delta U_{a} =-
{X_{a}}^b U_{b},
\end{equation}
y se introducen las derivadas covariantes asociadas a estas
transformaciones

\begin{eqnarray}\label{eq:ap219}
\mathcal{D}_{\mu}U_{a} & = & \partial_{\mu} U_{a} - {\omega_{\mu a}}^{b} U_{b}, \ \ \ \textrm{(a)} \nonumber\\
\mathcal{D}_{\mu}V^{a} & = & \partial_{\mu} V^{a} + V^{b}
{\omega_{\mu b}}^{a},\ \ \ \textrm{(b)}
\end{eqnarray}
con ${\omega_{\mu a}}^{b}$, la \emph{conexi\'on de spin}, que
transforma como

\begin{equation}
\delta {\omega_{\mu a}}^{b}=- \mathcal{D}_{\mu} {X_{a}}^{b}.
\end{equation}

Para objetos, mixtos como la tr\'iada, la derivada covariante total es

\begin{equation}
\mathcal{D}_{\mu}{e_{\nu}}^{a}=\mathcal{D}_{\mu}{e_{\nu}}^{a} -
\Gamma_{\mu\nu}^{\lambda}{e_{\lambda}}^{a}.
\end{equation}
As\'i al igual que pedimos que
$\mathcal{D}_{\mu}g_{\lambda\rho}=0$, si pedimos que
$\mathcal{D}_{\mu}{e_{\nu}}^{a}=0$ tendremos que

\begin{equation}
\mathcal{D}_{\mu}{e_{\nu}}^{a}=\Gamma_{\mu\nu}^{\lambda}{e_{\lambda}}^{a},
\end{equation}
de donde se obtiene que

\begin{eqnarray}
T_{\mu\nu}^{\lambda}{e_{\lambda}}^{a} & = & (\mathcal{D}_{\mu}{e_{\nu}}^{a} - \mathcal{D}_{\nu}{e_{\mu}}^{a}), \nonumber \\
& \equiv & {T_{\mu\nu}}^{a}.
\end{eqnarray}
Nos referiremos indistintamente a ${T_{\mu\nu}}^{a}$ como la
torsi\'on.

En $2+1$ dimensiones, dado que
${\omega_{\mu}}^{ab}=-{\omega_{\mu}}^{ba}$, introducimos la
conexi\'on de spin dual

\begin{equation}
{\omega_{\mu}}^{a}\equiv\frac{1}{2}\varepsilon^{abc}\omega_{\mu
bc}.
\end{equation}
De esto (\ref{eq:ap219}) se expresa como

\begin{eqnarray}
\mathcal{D}_{\mu}U_{a} & = & \partial_{\mu}U_{a} - {\varepsilon_{ab}}^{c} \omega_{\mu}^{b}U_{c}, \ \ \  \ \textrm{(a)}  \nonumber\\
\mathcal{D}_{\mu}V^{a} & = & \partial_{\mu}V^{a} -
{\varepsilon^{a}}_{bc}\omega_{\mu}^{b}V^{c}.\ \ \ \textrm{(b)}
\end{eqnarray}

El conmutador de derivadas covariantes nos permite introducir al an\'alogo del tensor de curvatura

\begin{equation}
[\mathcal{D}_{\mu},\mathcal{D}_{\nu}]V^{a}=V^{b}{R_{\mu\nu
b}}^{a},
\end{equation}
con

\begin{eqnarray}
{R_{\mu\nu b}}^{a}=\partial_{\mu}{\omega_{\nu b}}^{a} -
\partial_{\nu}{\omega_{\mu b}}^{a} + {\omega_{\nu a}}^{c}{\omega_{\mu c}}^{b} - {\omega_{\nu}}^{bc}\omega_{\mu
ca}.
\end{eqnarray}

Resulta que ${R_{\mu\nu}}^{ab}=-{R_{\mu\nu}}^{ba}$ y en $2+1$
introducimos

\begin{eqnarray}
{R_{\mu\nu}^{*}}^{a} & = & \frac{1}{2}{\varepsilon^{a}}_{bc}{R_{\mu\nu}}^{bc}, \nonumber \\
& = & \partial_{\mu}{\omega_{\nu}}^{a} -
\partial_{\nu}{\omega_{\mu}}^{a} -
{\varepsilon^{a}}_{bc}{\omega_{\mu}}^{b}{\omega_{\nu}}^{b}.
\end{eqnarray}
Tambi\'en sucede que ${R_{\mu\nu}^{*}}^{a}=-{R_{\nu\mu}^{*}}^{a}$,
as\'i que podemos introducir su dual

\begin{eqnarray}
R^{**\mu a} & = & \frac{1}{2}\varepsilon^{\mu\nu\lambda}{R_{\nu\lambda}^{*}}^{a}, \ \ \ \ \ \ \ \ \ \textrm{(a)}\nonumber \\
& = & \frac{1}{4}
\varepsilon^{\mu\nu\lambda}\varepsilon^{abc}R_{\nu\lambda bc}.\ \
\ \textrm{(b)}
\end{eqnarray}

Resulta que

\begin{eqnarray}
{R_{\mu\nu\lambda}}^{\sigma} & = & {R_{\mu\nu a}}^{b}{e_{\lambda}}^{a}{e^{\sigma}}_{b}, \\
R^{**\mu a} & = & -eG^{\mu\nu}{e_{\nu}}^{a},\label{eq:ap225}
\end{eqnarray}
con

\begin{equation}
 e=-\frac{1}{3!}
 \varepsilon_{abc}\varepsilon^{\mu\nu\lambda}{e_{\mu}}^{a}{e_{\nu}}^{b}{e_{\lambda}}^{c},
\end{equation}
el determinante de la tr\'iada. Las identidades de Bianchi que
surgen de las identidades de Jacobi para los conmutadores son

$$\mathcal{D}_{\mu}{R_{\nu\lambda}^{*}}^{a} + \ (\textrm{permutaciones c\'iclicas en}\  \mu\nu\lambda) = 0,$$
de donde

 $$\mathcal{D}_{\mu}R^{**\mu a}=0.$$

En otra direcci\'on la ecuaci\'on de ${T_{\mu\nu}}^{a}=0$ permite
despejar la conexi\'on en funci\'on de la tr\'iada

\begin{equation}\label{eq:ap227}
e{\omega_{\mu}}^{a}=(e_{\mu b}{e_{\rho}}^{a} -
\frac{1}{2}{e_{\mu}}^{a}e_{\rho
b})\varepsilon^{\rho\nu\lambda}\partial_{\nu}{e_{\lambda}}^{b}
\end{equation}

\section{Transformaciones de calibre en el lenguaje de las tr\'iadas}

La transformaci\'on de la conexi\'on de torsi\'on nula se relaciona con la  de la tr\'iada partiendo de (\ref{eq:ap227})

$$e\delta{\omega_{\mu}}^{a}=(e_{\mu b}{e_{\rho}}^{a} - \frac{1}{2}{e_{\mu}}^{a}e_{\rho b})
\varepsilon^{\rho\nu\lambda}\mathcal{D}_{\nu}\delta
{e_{\lambda}}^{b}.$$

Para los distintos tipos de invariancia tendremos

\emph{Lorentz:}

\begin{eqnarray}\label{eq:ap228}
\delta {e_{\mu}}^{a} & = & -{X^{a}}_{b}{e_{\mu}}^{b},\ \ \ \  \textrm{(a)} \nonumber \\
& \equiv & {\varepsilon^{a}}_{bc}l^{b}{e_{\mu}}^{c},\ \ \  \ \ \textrm{(b)} \\
\delta{\omega_{\mu}}^{a} & = & -\mathcal{D}_{\mu}l^{a}.\ \ \ \ \
\textrm{(c)}\nonumber
\end{eqnarray}

\emph{Difeomorfismo:}

\begin{eqnarray}
\delta {e_{\mu}}^{a} & = & \zeta^{\nu}\partial_{\nu}{e_{\mu}}^{a} + (\partial_{\mu}\zeta^{\nu}){e_{\nu}}^{a},\ \ \ \ \ \ \textrm{(a)} \nonumber \\
\delta{\omega_{\mu}}^{a} & = &
\zeta^{\nu}\partial_{\nu}{\omega_{\mu}}^{a} +
(\partial_{\mu}\zeta^{\nu}){\omega_{\nu}}^{a}.\ \ \ \  \
\textrm{(b)}
\end{eqnarray}
Estos \'ultimos pueden reescribirse como

\begin{eqnarray}
\delta {e_{\mu}}^{a} & = & \zeta^{\nu}{T_{\mu\nu}}^{a} + \mathcal{D}_{\nu}\zeta^{a} - {\varepsilon^{a}}_{bc}l_{\zeta}^{b}{e_{\mu}}^{c},\ \ \ \ \ \ \textrm{(a)} \nonumber \\
\delta{\omega_{\mu}}^{a} & = & -\mathcal{D}_{\nu}l_{\zeta}^{a} +
\zeta^{\nu}{R_{\mu\nu}^{*}}^{a}.\ \ \ \ \ \ \ \ \ \ \ \ \ \ \ \
\textrm{(b)}
\end{eqnarray}
con

\begin{eqnarray}
\zeta^{a} & = & \zeta^{\nu}{e_{\nu}}^{a},\ \ \ \ \ \ \textrm{(a)} \nonumber \\
l_{\zeta}^{a} & = & -\zeta^{\nu}{\omega_{\mu}}^{a}.\ \ \
\textrm{(b)}
\end{eqnarray}

\emph{Transformaciones ``conformes'':}

\begin{eqnarray}
\delta {e_{\mu}}^{a} & = & \frac{1}{2}\rho {e_{\mu}}^{a}\ \ \ \ \ \ \ \ \ \ \ \ \ \textrm{(a)} \nonumber \\
e\delta{\omega_{\mu}}^{a}  & = &
\frac{1}{2}{e_{\lambda}}^{a}{\varepsilon_{\mu}}^{\lambda\sigma}\partial_{\sigma}\rho.
\ \ \ \ \ \ \textrm{(b)}
\end{eqnarray}

\section{Objetos involucrados y definiciones en la formulaci\'on linealizada}

Linealizamos tomando

\begin{equation}
{e_{\mu}}^{a}=\delta_{\mu}^{a} + k{h_{\mu}}^{a} + O(k^{2}),
\end{equation}
as\'i $ e=1 + k{h_{\mu}}^{a}\delta^{\mu}_{a} + O(k^{2})$, y
observamos en (\ref{eq:ap227}) que ${\omega_{\mu}}^{a}$  es de
orden $k$. De hecho obtenemos

\begin{eqnarray}\label{eq:ap234}
{\omega_{\mu}^{L}}^{a}(h)& = &(\eta_{\mu b}\delta_{\rho}^{a} -
\frac{1}{2}\delta_{\mu}^{a}\eta_{\rho b})\varepsilon^{\rho\nu\lambda}\partial_{\nu}{h_{\lambda}}^b,\ \ \ \ \ \ \textrm{(a)} \nonumber\\
& \equiv &   {[{W_{\mu}}^{a}]^{\lambda}}_{b} {h_{\lambda}}^{b},  \ \ \ \ \ \ \ \ \ \ \ \ \ \ \ \ \ \ \  \ \ \ \ \  \ \textrm{(b)}\\
& \equiv & {W_{\mu}}^{a}(h).\ \ \ \ \ \ \ \ \ \ \ \ \ \ \ \ \ \ \
\ \ \ \ \ \  \ \ \ \ \ \textrm{(c)} \nonumber
\end{eqnarray}

(\ref{eq:ap234}a) puede escribirse, de forma mas sugestiva como,

\begin{equation}
{\omega_{\mu}^{L}}^{a}(h)=\frac{1}{2}\delta_{\lambda}^{a}\varepsilon^{\lambda\nu\gamma}[\partial_{\nu}(h_{\mu\gamma}
+h_{\gamma\mu}) -\partial_{\mu} h_{\nu\gamma}],
\end{equation}
Luego con (\ref{eq:ap225}) obtendremos

\begin{eqnarray}
G^{L\mu\nu}(h) & = & -\varepsilon^{\nu\rho\sigma}\partial_{\rho}{\omega_{\sigma}}^{c}(h)\delta^{\mu}_{c},\ \ \ \ \ \ \ \ \  \ \ \  \  \ \ \ \ \  \ \ \ \textrm{(a)} \nonumber \\
& = &  -\varepsilon^{\mu\alpha\rho}\varepsilon^{\nu\beta\sigma}\partial_{\alpha}\partial_{\beta}h_{\rho\sigma},\ \ \ \ \ \ \  \ \ \ \ \  \ \  \ \ \ \ \textrm{(b)}\\
& = & -
\frac{1}{2}\varepsilon^{\mu\alpha\rho}\varepsilon^{\nu\beta\sigma}\partial_{\alpha}\partial_{\beta}(h_{\rho\sigma}
+h_{\sigma\rho}),\ \ \ \ \ \ \textrm{(c)}\nonumber
\end{eqnarray}
de donde vemos que $G^{L\mu\nu}=G^{L\nu\mu}$ y que
$\partial_{\mu}G^{L\mu\nu}=0$. De la definici\'on de
$\widetilde{R}^{\mu\nu}$, \'este adquiere la forma

\begin{eqnarray}
\widetilde{R}^{L\mu\nu} & = & G^{L\nu\mu} - \frac{1}{2}\eta^{\mu\nu}G^{L},\nonumber \\
& = & -(\delta^{\mu}_{c}\delta^{\nu}_{\lambda} -
\frac{1}{2}\eta^{\mu\nu}\eta_{\lambda c})
\varepsilon^{\lambda\rho\sigma}\partial_{\rho}{\omega_{\sigma}}^{c}(h),
\end{eqnarray}
cuya forma recuerda a (\ref{eq:ap234}a). As\'i podremos
reescribirlo como

\begin{equation}
\widetilde{R}^{L\mu\nu}= -
\frac{1}{2}\varepsilon^{\nu\rho\sigma}[\partial_{\rho}(\omega_{\sigma}^{\mu}
+ \omega^{\mu}_{\sigma}) - \partial^{\mu}\omega_{\rho\sigma}].
\end{equation}
Por tanto el tensor de Cotton queda como

\begin{equation}
C^{L\mu\nu}=- \frac{1}{2}\varepsilon^{\mu\alpha\rho}\varepsilon^{\mu\beta\sigma}
\partial_{\alpha}\partial_{\beta}(\omega_{\rho\sigma}(h)
+\omega_{\sigma\rho}(h)).
\end{equation}

Esta similitud en las expresiones de
$\omega^{La}_{\mu}$,$G^{L\mu\nu}$ y $R^{L\mu\nu}$ en funci\'on de
${h_{\mu}}^{a}$ y ${\omega_{\mu}}^{a}$ va a ser importante cuando
miremos las realizaciones del \'algebra de Poincar\'e.

\section{Transformaciones a nivel linealizado}

Considerando el proceso de linealizaci\'on tendremos que para las
transformaciones bajo difeomorfismos infinitesimales

\begin{equation}
\delta_{\zeta}{h_{\mu}}^{a}=\partial_{\mu}\zeta^{a},
\end{equation}
y de (\ref{eq:ap234}) es claro que (asumimos ${T_{\mu\nu}}^{a}=0$)

\begin{equation}
\delta_{\zeta}\omega_{\mu}^{La}=0
\end{equation}

An\'alogamente para transformaciones infinitesimales de Lorentz sucede de (\ref{eq:ap228}) y (\ref{eq:ap234})

\begin{eqnarray}
\delta_{l}{h_{\mu}}^{a} & = & {\varepsilon^{a}}_{bc}l^{b}\delta_{\mu}^{c},    \nonumber \\
& = & -{\varepsilon_{\mu c}}^{a}l^{c},\ \ \ \ \ \ \textrm{(a)} \\
\delta_{l}\omega_{\mu}^{a} & = & -\partial_{\mu}l^{a}\ \ \ \ \ \ \textrm{(b)}\nonumber
\end{eqnarray}

As\'i cuando un objeto, como $G^{L\mu\nu}$, depende de solamente
de la parte sim\'etrica de ${h_{\mu}}^{a}$ diremos que es
expl\'icitamente invariante bajo transformaciones de Lorentz. De
hecho  $G^{L\mu\nu}$ es invarainte bajo los cambios

\begin{equation}
\delta h_{\mu\nu}=\delta_{\mu}\zeta_{\nu} -
\varepsilon_{\mu\nu\lambda}l^{\lambda},
\end{equation}
el pr\'imer t\'ermino es una transformaci\'on de difeomorfismo y
el segundo una transformaci\'on de Lorentz.

Finalmente para transformaciones conformes tendremos:

\begin{eqnarray}
\delta_{\rho}{h_{\mu}}^{a} & = & \frac{1}{2}\rho\delta_{\mu}^{a},\nonumber \\
\delta_{\rho}\omega_{\mu}^{La}  & = &
\frac{1}{2}{\varepsilon_{\mu}}^{\nu\sigma}\partial_{\sigma}\rho\delta_{\nu}^{a}.
\end{eqnarray}
Observamos que un objeto que dependa de la parte sim\'etrica de
$\omega_{\mu}^{La}$ es expl\'icitamente invariante bajo
transformaciones conformes, como es el caso del tensor de Cotton
linealizado.

\section{Descomposici\'on en partes irreducibles en la formulaci\'on linealizada}

Introduciendo los objetos

\begin{eqnarray}
\rho_{\mu} & = & \frac{\partial_{\mu}}{\Box^{1/2}}\ \ \ \ \ \  \ \ \ \  \ \ \textrm{(a)}\nonumber \\
\textrm{y} \ \ \ P^{\mu}_{\nu} & = & \delta^{\mu}_{\nu} - \rho^{\mu}\rho_{\nu}\ \ \ \ \ \ \textrm{(b)}
\end{eqnarray}

Descomponemos al tensor $h_{\mu\nu}$ (asociado a la
linealizaci\'on de la tr\'iada):

\begin{eqnarray}
h_{\mu\nu} & = & (H^{Tt}_{\mu\nu} + \frac{1}{2}P_{\mu\nu}H^{T} + \rho_{\mu}\rho_{\nu}H^{L} +
\rho_{\mu}h_{\nu}^{T} +\rho_{\nu}h_{\nu}^{T} )+ \nonumber \\
\qquad \qquad \qquad & &+(\varepsilon_{\mu\nu\lambda}V^{T\lambda}
+ \varepsilon_{\mu\nu\lambda}\rho^{\lambda}V^{L}),
\qquad \qquad \qquad \qquad \qquad \textrm{(a)} \\
& \equiv & h^{S}_{\mu\nu} + h^{A}_{\mu\nu},\ \ \ \ \ \  \ \ \ \ \
\ \ \  \ \ \ \ \ \ \  \ \ \ \ \   \qquad \qquad \qquad \ \ \
\textrm{(b)}\nonumber
\end{eqnarray}
donde

\begin{eqnarray}
h^{S}_{\mu\nu} & = & h^{S}_{\nu\mu},\ \ \ \ \ \  \ \ \ \ \ \  \ \  \  \ \ \ \ \ \ \ \  \  \ \ \ \ \ \ \ \textrm{(a)}\nonumber \\
h^{A}_{\mu\nu} & = & -h^{A}_{\nu\mu}, \ \ \ \ \ \  \ \ \ \ \ \  \ \ \ \ \ \ \ \ \  \ \ \ \ \ \ \ \ \textrm{(b)}\nonumber \\
H_{\mu}^{Tt\mu} & = & 0 \qquad, \qquad \partial_{\mu}H^{Tt\mu}_{\nu}=0,\ \ \ \ \ \ \textrm{(c)} \\
\partial_{\mu}h^{T\mu} & = & 0, \ \ \ \ \ \ \ \ \ \ \ \ \ \ \ \ \ \ \ \ \ \ \ \  \ \ \ \ \ \ \ \ \ \ \textrm{(d)}\nonumber \\
\partial_{\mu}V^{T\mu} & = & 0.\ \ \ \ \ \ \ \ \ \ \ \ \ \  \ \ \ \ \ \ \  \ \ \ \ \ \ \ \ \ \ \ \ \  \textrm{(e)}\nonumber \\
\end{eqnarray}

Una descompsici\'on an\'aloga se puede realizar sobre
${\omega_{\mu}}^{a}(=\eta^{a\nu}\omega_{\mu\nu})$

\begin{eqnarray}
\omega_{\mu\nu}& = & (W_{\mu\nu}^{Tt} + \frac{1}{2}P_{\mu\nu}W^{T} + \rho_{\mu}\rho_{\nu}W^{L} + \rho_{\mu}W^{T}_{\nu} + \rho_{\nu}W^{T}_{\mu})+\nonumber \\
\qquad \qquad \qquad & &
+(\varepsilon_{\mu\nu\lambda}a^{T\lambda}
+\varepsilon_{\mu\nu\lambda}\rho^{\lambda}a^{L}),\qquad \qquad  \qquad \qquad \textrm{(a)}\nonumber \\
& = & \omega_{\mu\nu}^{S} + \omega_{\mu\nu}^{A}.\ \ \ \ \ \ \ \ \
\ \ \ \ \ \  \ \ \ \qquad \qquad \qquad \qquad
\textrm{(b)}\nonumber
\end{eqnarray}

En funci\'on de las componentes irreducibles de $h_{\mu\nu}$ se tiene que

\begin{eqnarray}\label{eq:ap248}
W_{\mu\nu}^{Tt} & = & {\varepsilon_{\mu}}^{\sigma\lambda}\partial_{\sigma}H_{\lambda\sigma}^{Tt},\ \ \ \ \ \  \  \ \ \  \ \ \ \ \ \ \ \   \ \ \  \ \ \ \ \textrm{(a)}\nonumber \\
\omega_{\mu}^{T} & = & \frac{1}{2}{\varepsilon_{\mu}}^{\sigma\lambda}\partial_{\sigma}(h_{\lambda}^{T}
+ \varepsilon_{\lambda}^{\gamma\theta}\rho_{\gamma}V_{\theta}^{T}),\ \ \ \ \ \ \  \ \textrm{(b)}\nonumber \\
W^{T} & = & 0,\ \ \ \ \ \ \ \ \ \ \ \ \ \ \ \ \ \  \ \ \ \ \ \ \ \ \ \ \ \ \ \ \ \ \ \ \ \  \textrm{(c)} \\
W^{L} & = & \Box^{1/2}V^{L},\ \ \ \ \ \   \ \ \ \ \  \ \  \ \ \ \ \ \ \ \ \ \ \ \ \ \ \ \ \  \textrm{(d)}\nonumber \\
a_{\mu}^{T} & = & -\frac{1}{2}\Box^{1/2}(h_{\mu}^{T} +
{\varepsilon_{\mu}}^{\lambda\sigma} \rho_{\lambda}V_{\sigma}^{T}),\ \ \ \ \ \ \ \ \textrm{(e)}\nonumber \\
a^{L} & = & \frac{1}{2}\Box^{1/2}H^{T},\ \ \ \ \ \ \ \ \ \ \ \ \ \
\ \ \ \ \ \ \ \ \ \ \ \ \ \  \textrm{(f)}\nonumber
\end{eqnarray}

Bajo difeomorfismos sucede que $H_{\mu\nu}^{Tt}$, $H^{T}$ y
$V^{L}$ son invariantes, as\'i como  la combinaci\'on
$B_{\mu}^{T}=h_{\mu}^{T}
+{\varepsilon_{\mu}}^{\alpha\beta}\rho_{\alpha}V_{\beta}^{T}$. Se
ve entonces de la expresi\'on (\ref{eq:ap248}) que
$\delta_{\zeta}\omega_{\mu\nu}=0$.

Las transformaciones de Lorentz, como dijimos, s\'olo afectan la parte antisim\'etrica de $h_{\mu\nu}$ $(\delta_{l}V_{\mu}^{T}=-l_{\mu}^{T}, \delta V^{L}=-l^{L} ,\ \  \textrm{si} \ \  l_{\mu}=l_{\mu}^{T}+\rho_{\mu}l^{L} )$.

Para $\omega_{\mu\nu}$, las transformaciones de Lorentz solo afectan a $W_{\mu}^{T}$, $W^{L}$ y $a_{\mu}^{T}$

\begin{eqnarray}
\delta W_{\mu}^{T} & = & -\frac{1}{2} \Box^{1/2} l_{\mu}^{T},\ \ \ \ \ \ \ \ \ \ \ \ \ \  \textrm{(a)}\nonumber \\
\delta W^{L} & = & -\Box^{1/2} l^{L},\ \ \ \ \  \ \ \  \  \ \ \ \ \ \ \ \textrm{(b)}\\
\delta a_{\mu}^{T} & = & \frac{1}{2}{\varepsilon_{\mu}}^{\alpha\beta}\partial_{\alpha} l_{\beta}^{T},\ \ \ \ \ \ \ \ \ \ \ \ \ \ \textrm{(c)}\nonumber \\
\delta W_{\mu\nu}^{Tt} & = & \delta W^{T} = \delta a^{L}=0,\ \ \ \
\ \ \textrm{(d)}\nonumber
\end{eqnarray}
en analog\'ia con las transformaciones bajo difeomorfismos del $h_{\mu\nu}$.

La expresi\'on de $G^{L\mu\nu}$ en funci\'on de las partes
irreducibles de $h_{\mu\nu}$ es

\begin{equation}\label{eq:ap250}
G^{L}_{\mu\nu}= -\Box(H_{\mu\nu}^{Tt} -
\frac{1}{2}P_{\mu\nu}H^{T}),
\end{equation}
quedando expl\'icita la invariacia bajo difeomorfismos y
transformaciones de Lorentz del tensor de Einstein linealizado.

Para las transformaciones conformes sucede que

\begin{eqnarray}
\delta H_{\mu\nu}^{Tt} & = & \delta h_{\mu}^{T}= \delta h_{\mu\nu}^{a}=0 \ \ \ \  \ \ \ \  \ \ \ \ \textrm{(a)}\nonumber \\
\delta H^{T} & = & \rho \quad;\quad \delta H^{L}=\frac{1}{2}\rho \
\ \ \ \ \ \ \ \ \textrm{(b)}
\end{eqnarray}

\begin{eqnarray}
\delta a^{T} & = & - \frac{1}{2}\Box^{1/2}\rho,\ \ \ \  \textrm{(a)}\nonumber \\
\delta\omega_{\mu\nu}^{S} & = & \delta a_{\mu}^{T}=0.\ \ \ \ \ \
\textrm{(b)}
\end{eqnarray}

El tensor de Cotton puede verse que es expl\'icitamente invariante
bajo difeomorfirmos, transformaciones de Lorentz y conformes. Su
expresi\'on en funci\'on de las partes irreducibles de
$h_{\mu\nu}$ es

\begin{equation}\label{eq:ap253}
C^{\mu\nu}=-\Box\varepsilon^{\mu\alpha\beta}\partial_{\alpha}H_{\beta}^{Tt\nu}.
\end{equation}

\section{Tensores relevantes en funci\'on de solamente las componentes irreducibles de spin 2 }

En la descomposici\'on del  $h_{\mu\nu}$ distinguimos las componentes de spin 2 $H_{\mu\nu}^{Tt}$; las de spin 1: $h_{\mu}^{T}$, $V_{\mu}^{T}$; y las de spin 0: $H^{T}$, $H^{L}$, $V^{L}$. Una teor\'ia que describa a una part\'icula con spin 2 masiva debe proporcionar las restricciones din\'amicas para que las componentes asociadas a los spines menores $(1\ \textrm{y}\  0)$ se anulen. Siendo as\'i el caso se tendr\'a

\begin{equation}
h_{\mu\nu}=H_{\mu\nu}^{Tt},
\end{equation}
luego de (\ref{eq:ap248}),(\ref{eq:ap250}) y (\ref{eq:ap253})

\begin{eqnarray}
\omega_{\mu\nu}(h^{Tt}) & = & W_{\mu\nu}^{Tt}(h^{Tt}),\ \ \ \ \ \ \textrm{(a)}\nonumber \\
& = &
{\varepsilon_{\mu}}^{\sigma\lambda}\partial_{\sigma}H_{\lambda\nu}^{Tt},\
\ \ \ \ \ \textrm{(b)}
\end{eqnarray}

\begin{equation}
G_{\mu\nu}(h^{Tt})= - \Box H_{\mu\nu}^{Tt},
\end{equation}

\begin{eqnarray}
C_{\mu\nu}(h^{Tt}) & = & -\Box W^{Tt}(h^{Tt}),\ \ \ \ \ \ \textrm{(a)}\nonumber \\
& = & -\Box \varepsilon_{\mu}^{\sigma\lambda}\partial_{\sigma}
H_{\lambda\nu}^{Tt}.\ \ \ \ \ \ \textrm{(b)}
\end{eqnarray}
Estas \'ultimas relaciones se hacen presentes cuando miramos la
realizaci\'on de la condici\'on de Pauli-Lubansky para spin 2
donde, en el caso de la teor\'ia autodual, la variable fundamental
es $h_{\mu\nu}$, para la teor\'ia intermedia, es
$\omega_{\mu\nu}(h)$, y para la teor\'ia toplol\'ogica masiva
linealizada, es $G_{\mu\nu}(h)$.

\addcontentsline{toc}{chapter}{Bibliograf\'{\i}a}

\end{document}